\documentclass[ba,noinfoline]{imsart}
\usepackage{amsthm}
\usepackage{amsmath}
\usepackage{natbib}
\usepackage[colorlinks,citecolor=blue,urlcolor=blue,filecolor=blue,backref=page]{hyperref}
\usepackage{graphicx}
\usepackage{amssymb}

\startlocaldefs
\numberwithin{equation}{section}
\theoremstyle{plain}

\endlocaldefs

\theoremstyle{definition}
\newtheorem{exmp}{Example}[section]
\newcommand{\tr}{\text{tr}}

\makeatletter
\renewcommand{\ps@copyright}{%
  \renewcommand{\@evenhead}{\relax}%
  \renewcommand{\@oddhead}{\relax}%
  \renewcommand{\@evenfoot}{\relax}%
   \renewcommand{\@oddfoot}{\relax}}
\renewcommand{\ps@imsheadings}{%
  \renewcommand{\@evenhead}{\relax}%
  \renewcommand{\@oddhead}{\relax}%
  \renewcommand{\@evenfoot}{\relax}%
   \renewcommand{\@oddfoot}{\relax}}
\makeatother

\begin{document}

\begin{frontmatter}
\title{Using prior expansions for prior-data conflict checking}
\runtitle{Prior expansions}

\begin{aug}
\author{\fnms{David J.} \snm{Nott}\thanksref{addr1,addr2}\ead[label=e1]{standj@nus.edu.sg}},
\author{\fnms{Max} \snm{Seah}\thanksref{addr3}\ead[label=e2]{maxseah89@gmail.com}},
\author{\fnms{Luai} \snm{Al-Labadi}\thanksref{addr4}\ead[label=e3]{luai.allabadi@gmail.com}},
\author{\fnms{Michael} \snm{Evans}\thanksref{addr5}\ead[label=e4]{mevans@utstat.utoronto.ca}},
\author{\fnms{Hui Khoon} \snm{Ng}\thanksref{addr6,addr3,addr7}\ead[label=e5]{cqtnhk@nus.edu.sg}},
\author{\fnms{Berthold-Georg} \snm{Englert}\thanksref{addr3,addr7,addr8}\ead[label=e6]{cqtebg@nus.edu.sg}}

\runauthor{D. Nott et al.}

\address[addr1]{
  Department of Statistics and Applied Probability, National University of Singapore, Singapore 117546
  \printead{e1}
}

\address[addr2]{
  Institute of Operations Research and Analytics, National University of Singapore, Singapore 117546
}

\address[addr3]{
  Centre for Quantum Technologies, National University of Singapore, Singapore, 117543
  \printead{e2}
  \printead{e5}
  \printead{e6}
}

\address[addr4]{
  Department of Mathematical and Computational Sciences,
  University of Toronto Mississauga, Mississauga, Ontario LSL 1CG Canada
  \printead{e3}
}

\address[addr5]{
  Department of Statistics,
  University of Toronto, Toronto, ON M5S 3G3, Canada
  \printead{e4}
}
 
\address[addr6]{
  Yale-NUS College, Singapore 138614
}

\address[addr7]{
  MajuLab, CNRS-UNS-NUS-NTU International Joint Research Unit,
  UMI 3654, Singapore
}

\address[addr8]{
  Department of Physics, National University of Singapore, Singapore 117542
}

\end{aug}

\begin{abstract}
Any Bayesian analysis involves combining information represented through different model components, and when different
sources of information are in conflict it is important to detect this.  Here we consider checking for prior-data conflict in Bayesian models by expanding the prior used for the analysis
into a larger family of priors, and considering a marginal likelihood score statistic for the expansion parameter.  
Consideration of different expansions can be informative about the nature
of any conflict, and an appropriate choice of expansion can provide more sensitive checks for conflicts of certain types.  
Extensions to hierarchically specified priors and 
connections with other approaches to prior-data conflict checking are considered, and   
implementation in complex situations is illustrated with two applications.  The first concerns testing for the appropriateness of a LASSO
penalty in shrinkage estimation of coefficients in linear regression.  Our method is compared with a recent suggestion in the literature designed
to be powerful against alternatives in the exponential power family, and we use this family as the prior expansion for constructing our check.   
A second application concerns a problem in quantum state estimation, where a multinomial model is considered
with physical constraints on the model parameters.   In this example, the usefulness of different prior expansions is demonstrated for obtaining checks which are
sensitive to different aspects of the prior.
\end{abstract}


\end{frontmatter}

\section{Introduction}

A common approach to checking the likelihood in a statistical analysis is to consider model expansions motivated by
thinking about plausible departures from the assumed model.  
Then using either formal or informal methods for model choice,
we can compare the expanded model with the original one to determine whether the original model was good enough.  
In Bayesian analyses, information from the prior is combined with information in the likelihood, and an additional aspect 
of checking Bayesian models is to see whether the prior and likelihood information conflict. 
If the likelihood is inadequate, there will be no value of the model parameter giving a good fit to the data, whereas  prior-data conflict occurs
when the prior is putting all its mass out in the tails of the likelihood.  Checking for prior-data conflict is important, because it is not sensible to
combine conflicting sources of information without careful thought, and the influence of the prior only increases with increasing conflict.   

The purpose of this work is to consider model expansion for checking for prior-data conflict, rather than for checking the
likelihood.  Previous work on the use of model expansions for 
exploring structural model uncertainty such as in \cite{draper95} does not deal specifically with prior expansions or their
use for prior-data conflict checking.  
Existing checks for prior-data conflict are determined once the parameter to be checked and any hierarchical
structure of the prior is specified.  The method we suggest here is different, because the choice
of a particular prior expansion provides the flexibility to design checks which are sensitive to conflicts of
certain kinds.

Suppose that $\theta$ is a parameter, $y$ is data, $g(\theta)$ is a prior
density for $\theta$ and $p(y|\theta)$ is the sampling model. Write
$g(\theta|y)\propto g(\theta)p(y|\theta)$ for the posterior density. Suppose
that we have checked the likelihood component of the model and that it is
thought to be adequate, so that checking for prior-data conflict is of
interest. Checking the likelihood component first is important, since sound
inferences cannot result from a poor model no matter what prior is used for
$\theta.$  
As noted in \cite{allabadi+e17}, the existence of prior-data conflict
may be associated with sensitivity to the prior but, even if a conflict
exists, with sufficient data, the effect of the prior can be minimal. Such a
situation does not imply, however, that prior-data conflict is no longer of
interest. For if the prior was elicited, as it should be, 
then the existence of a prior-data conflict is informing the
participants of a problem with that procedure. Thus there is a need for
methods to assess prior-data conflict, and the need for formal procedures
is particularly apparent in multi-parameter settings where simple plots will not
suffice for this task. The developments in this paper are concerned with
providing suitable methodology for this problem.

Our approach to prior-data conflict checking considers embedding the original
prior into a larger family, which we write as $g(\theta|\gamma)$, where
$\gamma$ is some expansion parameter and the original prior is $g(\theta
|\gamma_{0})$ for some value $\gamma_{0}$. The corresponding posterior
distributions will be denoted by $g(\theta|y,\gamma)$. Throughout this work
$\gamma$ will be a scalar, or if we embed the prior into a family with more
than one additional parameter we will vary these parameters one by one. If we
integrate out $\theta$ from the likelihood using $g(\theta|\gamma)$, we
obtain
\begin{align*}
p(y|\gamma) & =\int g(\theta|\gamma)p(y|\theta)d\theta,
\end{align*}
and we propose using the score type statistic
\begin{align}
S(y) & =\left.  \frac{d}{d\gamma}\log p(y|\gamma)\right\vert _{\gamma=\gamma_{0}} \label{scorestat}
\end{align}
for checking for prior-data conflict. A $p$-value for the statistic
(\ref{scorestat}) is computed to provide a calibration of its value by using
the prior predictive density for the data to obtain the reference
distribution. This gives the $p$-value
\begin{align}
p_{S} & =P(S(Y)\geq S(y_{\mathtt{obs}})), \label{scorecheck}
\end{align}
where $Y\sim p(y)=\int g(\theta)p(y|\theta)d\theta$ and $y_{\mathtt{obs}}$ is
the observed value of $y$. In the next section we will describe a framework
for Bayesian model checking that explains some logical requirements that a
prior-data conflict check should satisfy, and we discuss why (\ref{scorecheck}) 
satisfies these requirements.

We show later that under appropriate regularity conditions (\ref{scorestat}) has the alternative expression
\begin{align}
  S(y) & = \int \left.\frac{d}{d\gamma} \log g(\theta |\gamma)\right|_{\gamma=\gamma_0} g(\theta|y)d\theta.  \label{scorestat2}
\end{align}
Expression (\ref{scorestat2}) gives an intuitive meaning to the test statistic (\ref{scorestat}), as well as being useful later for computation.  
We can see that (\ref{scorestat}) is the posterior expectation of the rate of change of the log prior with respect to the expansion parameter.  If there is a conflict,
and our posterior distribution is concentrated in the tails of the prior, then the derivative of the log prior with respect to the expansion parameter will be large if the prior
is changing in a direction that reduces the conflict when we vary $\gamma$ around $\gamma_0$.  
In Section 2.3 we give further motivation for a check based on (\ref{scorestat}) 
when we describe the relationship between the score
statistic and checks based on relative belief measures.
One of the main advantages of the check we propose is that by appropriate choices of the prior expansion we can obtain checks of conflict which are
sensitive to different aspects of the prior.  Other proposals for prior-data conflict checking do not have this feature, which will be illustrated in some of the examples.   

To avoid confusion we emphasize that 
in our work the parameter $\gamma$ is not considered a hyperparameter to be chosen
by an elicitation procedure.  Instead, 
the role of the family of priors 
$g(\theta|\gamma)$ is to assess whether the elicited prior $g(\theta)$
conflicts with the data when a certain discrepancy, derived from 
the chosen family, is being checked.  More than one family of priors and hence more than
one discrepancy may be used, and it is possible that no prior in a certain
prior family will pass all checks considered.  If $\gamma$ were to be chosen via an
elicitation to obtain a new prior, this new prior would also need to be checked for
conflict with the data.

The developments here can be seen as contributing to one aspect of a
much broader problem, namely, the sensitivity of a Bayesian statistical
analysis to the inputs chosen. Robustness to the
prior is a valid goal and, while the lack of prior-data conflict may give some
comfort in that regard, it cannot be claimed that this guarantees a lack of
sensitivity to the choice made. There is also the sensitivity to the choice of
the model and this also needs to be assessed. There is an extensive literature
on the general topic of Bayesian sensitivity analysis as found, for
example, in \cite{lavine91}, \cite{clarke+g98}, \cite{gustafson+c04}, 
\cite{zhu+it11} and \cite{roos+mhr15}.

In the next section, we discuss prior-data conflict checking and how this differs from checking the likelihood component in a model-based statistical analysis.  
We also review the existing literature on checking for prior-data conflict.  
Section 3 discusses hierarchical extensions of our check.  Section 4 illustrates implementation of the checks in two complex applications.
The first concerns shrinkage estimation of coefficients in linear regression with squared error loss and 
a LASSO penalty \cite{tibshirani96}, which can be thought of equivalently as MAP estimation
for a Gaussian linear regression model with a Laplace prior on the coefficients.  
\cite{griffin+h17} have recently considered a test for the appropriateness of the LASSO penalty based on the empirical
kurtosis for a point estimate of the coefficients, and where their test is designed to be powerful against alternative priors in an exponential power family.  
Here we consider the embedding of the Laplace prior into this same family for the construction of our prior-data conflict score test, and show that 
our method is an attractive one in this example.  
The second application considered relates to a problem in quantum tomography.  Here the model is a multinomial, but with physical constraints on the parameter space. 
We consider several different prior expansions leading to statistics that are sensitive to different aspects of the prior.  Section 5 gives some concluding
discussion.    

\section{Prior-data conflict checking}

In this section we explain the divergence-based checks
considered in \cite{nott+wee16}, and their connections with the score-based checks suggested here.
We follow this with a review of the wider prior-data conflict checking literature.  

\subsection{Conflict checks based on relative belief}

Prior-data conflict checks assess whether the prior puts all its mass out in the tails of the likelihood.  
Said another way, we want to see if the observed likelihood is surprising compared with what is expected for data
generated under the prior.  Hence similar to (\ref{scorestat}) and (\ref{scorecheck}), a prior-data
conflict check is a prior predictive check \cite{box80} which 
defines some statistic and compare its observed value to a reference distribution obtained
from the prior predictve density for the data $p(y)$.   \cite{evans+m06} observe that any statistic used for prior-data conflict
checking should depend on the data only through a minimal sufficient statistic $T$.  Since a minimal sufficient statistic
determines the likelihood, dependence of the statistic on other aspects of the data apart from $T$ 
is undesirable, since prior-data conflict has nothing
to do with aspects of the data irrelevant to the likelihood.  Based on this reasoning, 
and writing $p(t)=\int p(t|\theta)g(\theta)\,d\theta$ for the prior predictive density of $T$, \cite{evans+m06} suggested using the prior predictive $p$-value
\begin{align}
  p_{\text{EM}}& =P(p(T)\leq p(t_{\text{obs}}))  \label{emcheck}
\end{align}
to check for conflict, where $T\sim p(t)$ is a sample from the prior predictive for the minimal sufficient statistic $T$ and $t_{\text{obs}}$ is the observed value.  

\cite{evans+j10} note that (\ref{emcheck}) is not invariant to the minimal sufficient statistic chosen, and suggest an invariantized version of the check.  
\cite{nott+wee16} considered an alternative check based on prior to posterior R\'{e}nyi divergences \cite{renyi61} which is also invariant.  We describe this approach in detail, since it is closely related to our proposed score checks and gives
some additional motivation for them.  For the method of \cite{nott+wee16}, a conflict $p$-value is computed as
\begin{align}
  p_{\alpha} & = P(R_\alpha(Y)\geq R_\alpha(y_{\text{obs}}))  \label{klcheck}
\end{align}
where $Y$ denotes a draw from the prior predictive distribution for $y$, $p(y)=\int g(\theta)p(y|\theta)d\theta$, and $R_\alpha(y)$ denotes the prior
to posterior R\'{e}nyi divergence for data $y$, 
\begin{align*}
  R_\alpha(y)=\frac{1}{\alpha-1}\log \int \left\{\frac{g(\theta|y)}{g(\theta)}\right\}^{\alpha-1}g(\theta|y)d\theta,
\end{align*}
where $\alpha>0$ and the case $\alpha=1$ is defined by taking a limit $\alpha\rightarrow 1$, which corresponds to the Kullback-Leibler divergence.  

\cite{nott+wee16} note connections between their suggested check and
 the relative belief framework for inference \cite{evans15,baskurt+e13}.  For a parameter of interest
$\psi(\theta)$, the relative belief function for $\psi$ is 
$$\mbox{RB}(\psi|y)=\frac{g(\psi|y)}{g(\psi)},$$
where $g(\psi|y)$ is the posterior distribution for $\psi$, and $g(\psi)$ is the prior.  If the relative belief is larger than $1$ at a given $\psi$, this means 
there is evidence for that value, whereas if it is less than $1$ there is evidence against.  
$R_\alpha(y_{\text{obs}})$ is a measure of the average evidence in $y_{\text{obs}}$ or equivalently the average change in beliefs
from {\it a priori} to {\it a posteriori}.  
So (\ref{klcheck}) is a measure of how much beliefs about $\theta$ have changed from prior to posterior
compared with what is expected under the prior, and is hence a measure of how surprising the data are under the prior.  
Relative belief inferences have been shown to possess optimal robustness to the prior properties but this
robustness decreases with increasing prior data conflict, see \cite{allabadi+e17}.
The case $\alpha=2$ gives the posterior mean of the relative belief function, whereas
$\alpha\rightarrow\infty$ corresponds to the maximum relative belief. 

Because the discrepancy for the check $R_\alpha(y)$ depends on the data only through the posterior, this discrepancy is a
function of any minimal sufficient statistic, and it is invariant to the choice of sufficient statistic.  
There is also a connection between the check
(\ref{klcheck}) and the Jeffreys' prior.  \cite{nott+wee16} show that the limiting form of the $p$-value is
\begin{align}
  & P\left(g(\theta^*)|I(\theta^*)|^{-1/2} \geq g(\theta)|I(\theta)|^{-1/2}\right) \label{klchecklim}
\end{align}
where $\theta^*$ denotes the true parameter and $\theta\sim g(\theta)$.  The $p$-value (\ref{klchecklim}) is a measure of how far out in the tails of the prior density the true
parameter is (with the prior expressed as a density with respect to the Jeffreys' prior as support measure). 
A similar limiting result holds for the check
of \cite{evans+m06}, but where the prior density is expressed with respect to the Lebesgue measure as support measure.  

\subsection{Connections between the relative belief and score checks}

The score based statistic (\ref{scorestat}) is closely related to the divergence based check of \cite{nott+wee16}.  
First, it shares the property 
with (\ref{klcheck}) of depending on the data only through the posterior. 
This follows from the expression (\ref{scorestat2}) for $S(y)$, an expression which is derived from Fisher's identity (see, for example, \cite{cappe+mr05}, equation (10.12)).  
Fisher's identity applies when we have some model for data $y$, with latent variables $z$ and a parameter $\eta$.  There is a joint model for $(y,z)$ given $\eta$, 
$p(y,z|\eta)$ say, and $p(y|\eta)$ is obtained by integrating out the latent variables, 
$p(y|\eta)=\int p(y,z|\eta) dz$.  Fisher's identity states that under appropriate regularity conditions
\begin{align*}
\nabla_\eta \log p(y|\eta) & = \int \left(\nabla_\eta \log p(y,z|\eta)\right) p(z|y,\eta) dz.
\end{align*}
Using this formula and identifying $\theta$ with $z$ and $\gamma$ with $\eta$, the expression (\ref{scorestat2}) for $S(y)$ follows, provided
that the differentiation under the integral sign required for Fisher's identity is valid.  
Because the score check depends on the data only through the posterior, the statistic $S(y)$ depends only
on the data through the value of a minimal sufficient statistic, and it is invariant to the choice of that statistic.  This is desirable for a prior-data conflict check as discussed in Section 2.1.  Furthermore, to apply the method it is not required
to identify any non-trivial minimal sufficient statistic, since 
$S(y)$ is computed directly from the posterior distribution.

As well as depending on the data only through the posterior, the score based check has a motivation related to relative belief based inference.  
By rearranging Bayes' rule with the prior $g(\theta|\gamma)$, 
$$p(y|\gamma)=\frac{g(\theta|\gamma)p(y|\theta)}{g(\theta|y,\gamma)},$$
so that
\begin{align}
 \frac{d}{d\gamma} \log p(y|\gamma) & =-\frac{d}{d\gamma} \log \frac{g(\theta|y,\gamma)}{g(\theta|\gamma)}. \label{RBrep}
\end{align}
Hence $S(y)$ is the derivative with respect to the expansion parameter at $\gamma_0$ of the negative log relative belief, evaluated at any $\theta$.  
The right-hand side does not depend on $\theta$, and we can average over any distribution on $\theta$.  Averaging over $g(\theta|y,\gamma)$,
$$\frac{d}{d\gamma} \log p(y|\gamma)=-\frac{d}{d\gamma} \int \log \frac{g(\theta|y,\gamma)}{g(\theta|\gamma)} g(\theta|y,\gamma)d\theta.$$
Hence the score-based check statistic is the negative of the derivative with respect to $\gamma$ at $\gamma_0$ of the Kullback-Leibler divergence
statistic of \cite{nott+wee16}.  From (\ref{RBrep}) we see that if there are values $\theta$ where the posterior is large but the prior is small, which happens when there is a conflict, then if the prior value changes rapidly with respect to $\gamma$ then $\log p(y|\gamma)$ will also change rapidly.  This provides additional intuition for our score statistic.

A further connection between the score and relative belief approaches emerges by considering the expansion 
$g(\theta|\gamma)=(1-\gamma)g(\theta)+\gamma q(\theta)$ for a fixed prior $q(\theta)$.  Using (\ref{scorestat2}),
\begin{align*}
  S(y) & = E\left(\frac{q(\theta)-g(\theta)}{g(\theta)}\Big|y\right)  \\
 & = E\left(\frac{q(\theta)}{g(\theta)}\Big|y\right)-1. 
\end{align*}
If the Jeffreys' prior is proper, then taking $q(\theta)$ to be the Jeffreys' prior, (\ref{scorecheck}) becomes
\begin{align*}
  p_S & = P\left(E\left(\frac{1}{g(\theta)|I(\theta)|^{-1/2}}\Big|y_{\text{obs}}\right)\leq E\left(\frac{1}{g(\theta)|I(\theta)|^{-1/2}}\Big|Y\right)\right),
\end{align*}
for $Y\sim p(y)$,
and in the asymptotic limit this is equivalent to the $p$-value (\ref{klchecklim}) obtained using the divergence based check.  

\subsection{Other approaches to prior-data conflict checking}

There is an extensive existing literature on prior-data conflict checking.  
One class of approaches involves converting the likelihood and prior information into something comparable, either through renormalization or converting the likelihood
to a posterior through a non-informative prior.  A recent example of this approach is \cite{presanis+osd13}, where conflicts are examined locally at any node
or group of nodes in a directed acyclic graph.  Their work unifies and generalizes a number of previous suggestions
\cite{ohagan03,dahl+gn07,marshall+s07,gasemyr+n09}.   \cite{scheel+gr11}  consider a related approach where the model is formulated as a chain graph
and at a certain node a marginal posterior distribution based on a local prior and lifted likelihood are compared. \cite{bousquet08} considers an approach 
where ratios of prior-to-posterior Kullback-Leibler divergences are calculated, for the prior to be checked and a non-informative prior.  Hierarchical extensions are also discussed.  
\cite{reimherr+mn14} consider the difference in information required to be put into a likelihood function to obtain the same posterior uncertainty for a proper prior
used in an analysis relative to a non-informative baseline prior that would be used if little prior information were available.
Another method similar to those of \cite{evans+m06} and \cite{nott+wee16}, in not requiring 
the use of any non-informative prior, is described in \cite{dey+gsv98}, where vectors of quantiles of the posterior distribution itself are used in a Monte Carlo test using a prior predictive reference distribution.  
\cite{bayarri+c07} review and evaluate various methods for checking the second level of hierarchical models, and advocate
the partial posterior predictive $p$-value approach \cite{bayarri+b00}.  
General discussions of Bayesian model checking which are not specifically concerned with
checking for prior-data conflict are given in \cite{gelman+ms96}, \cite{bayarri+b00} and \cite{evans15}.
The method we propose here is a useful addition to the above proposals because the use of an appropriate encompassing family of priors for constructing the check
gives some guidance for how to construct checks that are sensitive to conflicts in different aspects of the prior;  furthermore, it does not rely on the construction
of any non-informative prior for its application, which can sometimes be difficult.  

There is also a growing literature on the question of what to do in a Bayesian analysis if a prior-data conflict is found.  It may
seem problematic from a Bayesian point of view to change the prior after looking at the data.  
We believe that how to handle a conflict depends on why the conflict occurred.  As an example, suppose that the
prior was formulated based on data from a previous experiment.  A Bayesian analysis is performed and a prior-data conflict
is detected.  Following this, further investigation showed that the data from the previous experiment was misreported.  
A new prior is then formulated based on the corrected data.
We think it is clear in this setting that although looking at the data resulted in the change of prior, 
the analysis based on the new prior does not have any problematic interpretation from a Bayesian point of view. Of course
not all cases are as clear cut as this one.

We do not really feel that checking the prior is different to other forms of Bayesian model checking focusing on the likelihood
in terms of needing to justify a data-driven change in the model.  Responding to a conflict requires judgements 
about how the deficiencies uncovered through model checking relate to prior information that was not used in the original 
elicitation of prior and model, based on limited time and thought.  
Another consideration in responding to a conflict is whether we need to do anything at all, 
since sometimes the data swamps the prior.  However, detecting conflicts in such a setting is still important
because it may reveal a defect in our understanding in setting up the model.  

One approach to modifying a prior when a conflict is detected is described in \cite{evans+j11}.  A definition is provided
there for what it means for a prior to be weakly informative with respect to another base prior.  The base
prior can be considered as the initial prior one would like to use in an analysis.  The definition of weak informativity
is then in terms of potential prior-data conflicts that one could encounter using the new prior and is quantitative
in the sense that a prior may lead to 50\% fewer prior-data conflicts than the base prior.  A hierarchy of progressively more
weakly informative priors can then be defined and this is done before seeing the data.  As such, if a prior-data conflict is
encountered, one can proceed up the hierarchy of priors until a conflict is avoided.  There is still a dependence
of the prior on  data, in the sense that a replacement prior is required, but the dependence is very weak, and
surely much less than what is encountered in the use of empirical Bayes methdology.  In essence, one prepares for
the possibility of prior-data conflict before seeing the data, and the hierarchy is part of the ingredients that go into an 
analysis.  This aspect of prior-data conflict is not pursued further here as our focus is on a new technique
for detecting conflicts.  See
\cite{evans+m06}, \cite{evans+j11}, \cite{held+s17} and \cite{bickel18} for additional perspectives.  

\section{Hierarchical extension of the score based check}

When a prior distribution is elicited hierarchically, it is desirable to check the different parts of the prior separately
since this can be more informative about what parts of the prior are problematic.  We describe how to
do this with the proposed score based checks.  Other methods in the literature
for checking for conflict at nodes of a graphical
model can also be used for checking hierarchical priors.  
Similar to the non-hierarchical case, our method is different to existing methods through allowing expansions
of different prior components allowing design of checks sensitive to different types of conflict.

Let $\theta$ be partitioned as $\theta=(\theta_1^\top,\theta_2^\top)^\top$ and suppose
the prior has been specified as $g(\theta)=g(\theta_1)g(\theta_2|\theta_1)$.  The discussion can be generalized to the case where $\theta$ is partitioned into
more than two parts.  First, consider an expansion of the form 
$g(\theta|\gamma^{(1)})=g(\theta_1)g(\theta_2|\theta_1,\gamma^{(1)})$, where the marginal prior for $\theta_1$ is held fixed but the conditional prior
$g(\theta_2|\theta_1)$ is embedded into $g(\theta_2|\theta_1,\gamma^{(1)})$ with $g(\theta_2|\theta_1,\gamma_0^{(1)})=g(\theta_2|\theta_1)$.  
Consider
\begin{align*}
p(y|\theta_1,\gamma^{(1)})=\int p(y|\theta)g(\theta_2|\theta_1,\gamma^{(1)}) d\theta_2,
\end{align*}
and define
\begin{align}
  S^{(1)}(y,\theta_1) & = \left.\frac{d}{d\gamma^{(1)}} \log p(y|\theta_1,\gamma^{(1)})\right|_{\gamma^{(1)}=\gamma_0^{(1)}}, \label{stat1}
\end{align}
and
\begin{align}
  S^{(1)}(y) & = E\left(S^{(1)}(y,\theta_1)|y_{\text{obs}}\right). \label{stat2}
\end{align}
We propose to check for conflict for the conditional prior $g(\theta_2|\theta_1)$ using the $p$-value
\begin{align*}
 p_{S1} & = P(S^{(1)}(Y)\geq S^{(1)}(y_{\text{obs}})),
\end{align*}
where $Y\sim m(y)=\int g(\theta_2|\theta_1)g(\theta_1|y_{\text{obs}})p(y|\theta) d\theta$.  Here it has been assumed again in the calculation of the $p$-value that the 
embedding prior family is such that a large value of $S^{(1)}(y)$ indicates conflict.  

The justification for this check is that in checking $g(\theta_2|\theta_1)$ we should consider an appropriate check for this prior as if $\theta_1$ is fixed (which leads
to the statistic $S^{(1)}(y,\theta_1)$) but then to eliminate the unknown $\theta_1$ we take the expectation with respect to $\theta_1$ under the posterior
given $y_{\text{obs}}$.  So we see if there is a conflict involving $g(\theta_2|\theta_1)$ for $\theta_1$ values that reflect knowledge of $\theta_1$ under $y_{\text{obs}}$.  
The reference distribution for the check also reflects knowledge of $\theta_1$ under $y_{\text{obs}}$ but using the conditional prior of $\theta_2$ given $\theta_1$ in
generating predictive replicates.  

To check $g(\theta_1)$, we now consider a different expansion $g(\theta_1|\gamma^{(2)})g(\theta_2|\theta_1)$ of the joint prior, where $g(\theta_1|\gamma_0^{(2)})=g(\theta_1)$ 
and then with $p(y|\gamma^{(2)})=\int p(y|\theta)g(\theta_1|\gamma^{(2)})g(\theta_2|\theta_1) d\theta$ consider the statistic
\begin{align*}
  S^{(2)}(y) & = \left. \frac{d}{d\gamma^{(2)}} \log p(y|\gamma^{(2)})\right|_{\gamma^{(2)}=\gamma_0^{(2)}},
\end{align*}
and a $p$-value for the check of $g(\theta_2)$ 
\begin{align*}
  p_{S2} & = P(S^{(2)}(Y) \geq S^{(2)}(y_{\text{obs}})),
\end{align*}
with $Y\sim m(y)=\int p(y|\theta)g(\theta) d\theta$.  It is again assumed in the $p$-value computation that the embedding prior family is such that large $S^{(2)}(y)$ indicates conflict.  

These checks are similar to those considered in \cite{nott+wee16} for their divergence based check. As explained there, in models with particular additional structure,
the checks above can be modified in various ways.  For example, in hierarchical models with observation or cluster specific parameters, cross-validatory versions of the check can be 
considered, as well as versions of partial posterior predictive checks \cite{bayarri+b00} in constructing $S^{(1)}(y)$ and its reference distribution.  If there are sufficient or ancillary statistics at different levels this can be exploited also \cite{evans+m06,nott+wee16}.  

\section{Simple examples}

It is insightful to consider properties of the check (\ref{scorecheck}) first in some simple examples, where calculations can be performed analytically.  To obtain tractable calculations, the examples are restricted to
exponential family models, and the prior expansions we use 
involve varying hyperparameters in conjugate priors.  
The expansions based on conjugate forms do not illustrate well the flexibility of our method, since there is no
freedom to choose the expansion used.  The 
more complex examples of Section 5 are more informative in this respect.  
However, the examples below are still interesting,
since our checks correspond to some of the standard ones in the literature for these cases.
The examples discussed were given in \cite{evans+m06} and \cite{nott+wee16}.  
We will use the following notation, which was also used in \cite{nott+wee16}.  If $S_1(y)$ and $S_2(y)$ are two discrepancies
for a Bayesian model check, and one is a monotone function of the other (as a function of $y$), we will write $S_1(y)\doteq S_2(y)$.  Note that prior predictive checks based
on these discrepancies will give the same result, if the appropriate tail probability is calculated.  

\begin{exmp}
{\it Normal location model}.  

\noindent
Let $y_1,\dots, y_n$ be a random sample, $y_i\sim N(\theta,\sigma^2)$, where $\sigma^2>0$ is a known variance and $\theta$ is an unknown mean.  
The sample mean is sufficient for $\theta$ and normally distributed, so it suffices to consider the case $n=1$ and this will be assumed in
what follows.  We write $y_{\text{obs}}$ for the observed value of $y$.  
Suppose the prior for $\theta$ is normal, $N(\mu_0,\tau_0^2)$, where $\mu_0$ and $\tau_0^2$ are fixed hyperparameters.  
Next, expand the prior to $N(\mu_0,\tau^2)$, where $\tau^2$ is allowed to vary.  
Clearly $p(y|\tau^2)$ is a normal density, with mean $\mu_0$ and variance $\sigma^2+\tau^2$, and hence
\begin{align*}
 \log p(y|\tau^2) & = -\frac{1}{2}\log2\pi(\sigma^2+\tau^2)-\frac{(y-\mu_0)^2}{2(\sigma^2+\tau^2)},
\end{align*}
which gives
\begin{align*}
  \left.\frac{d}{d\tau^2}\log p(y|\tau^2)\right|_{\tau^2=\tau_0^2} & \doteq \frac{(y-\mu_0)^2}{2(\sigma^2+\tau_0^2)^2}\doteq (y-\mu_0)^2.
\end{align*}
So to calculate the prior predictive $p$-value for the check we compare $(y_{\text{obs}}-\mu_0)^2$ to its prior predictive density.  This is the
same check obtained by \cite{evans+m06} and \cite{nott+wee16} using 
(\ref{emcheck}) and (\ref{klcheck}) and the corresponding $p$-value is (\cite{evans+m06}, p. 897)
$$2\left(1-\Phi\left(\frac{|y_{\text{obs}}-\mu_0|}{\sqrt{\sigma^2+\tau_0^2}}\right)\right).$$
\end{exmp}

\begin{exmp}
{\it Binomial model}

\noindent
Let $y\sim \mbox{Binomial}(n,\theta)$ where $\theta$ is unknown with a prior $g(\theta)$ that is a beta distribution, $\mbox{Beta}(a,b)$.  
The expansion we consider here is a geometric mixture of 
the $\mbox{Beta}(a,b)$ prior and the Jeffreys' prior, which is $\mbox{Beta}(1/2,1/2)$.  We denote the mixing parameter by $\gamma$ and
\begin{align*}
g(\theta|\gamma) & \propto \left\{\theta^{(a-1)}(1-\theta)^{(b-1)}\right\}^\gamma \times \left\{\theta^{-1/2}(1-\theta)^{-1/2}\right\}^{1-\gamma},
\end{align*}
which is a beta prior, $g(\theta|\gamma)=\mbox{Beta}(\gamma a+(1-\gamma)/2,\gamma b+(1-\gamma)/2)$.  
Hence $p(y|\gamma)$ is a beta-binomial probability function, 
\begin{align*}
 p(y|\gamma) & = \binom{n}{y}\frac{B(y+\gamma a+(1-\gamma)/2,n-y+\gamma b+(1-\gamma)/2)}{B(\gamma a+(1-\gamma)/2,\gamma b+(1-\gamma)/2)}.
\end{align*}
Taking logs and differentiating with respect to $\gamma$, we obtain
\begin{align*}
\frac{d}{d\gamma} \log p(y|\gamma)  & \doteq \left\{\psi\left(y+\gamma a+\frac{1-\gamma}{2}\right)-\psi\bigg(n+\gamma (a+b)+(1-\gamma)\bigg)\right\}(a-1/2) \\
 & \;\;\; +\left\{\psi\left(n-y+\gamma b+\frac{1-\gamma}{2}\right)-\psi\bigg(n+\gamma (a+b)+1-\gamma\bigg)\right\}(b-1/2),
\end{align*}
where $\psi(\cdot)$ denotes the digamma function.  Using the fact that $\psi(x)=\log x +O(1/x)$, we can write
\begin{align}
\left.\frac{d}{d\gamma} \log p(y|\gamma)\right|_{\gamma=1} & \doteq (a-1/2)\log \tilde{\theta}_n+(b-1/2)\log(1-\tilde{\theta}_n)+O\left(\frac{1}{n}\right), \label{binomial}
\end{align}
where $\tilde{\theta}_n=(y+a)/(n+a+b)$ is the posterior mean of $\theta$ under the prior $g(\theta)$.  
Equation (\ref{binomial}) is equivalent to
\begin{align}
\left.\frac{d}{d\gamma} \log p(y|\gamma)\right|_{\gamma=1} & \doteq \log g(\tilde{\theta}_n)+\frac{1}{2}\log |I(\tilde{\theta}_n)|+O\left(\frac{1}{n}\right),
\end{align}
where $I(\theta)=n/(\theta (1-\theta))$ is the Fisher information.  This matches the asymptotic form of the check (\ref{klcheck}) considered
in Section 4 of \cite{nott+wee16}.  We have already established in Section 3 that an arithmetic mixture involving the Jeffreys' prior would lead to
a similar result.  On the other hand, if instead of considering a geometric mixture of the Jeffreys' prior with the $\mbox{Beta}(a,b)$ prior we
instead consider a geometric mixture of the uniform distribution with $\mbox{Beta}(a,b)$ instead, then we obtain, using a similar argument,  
\begin{align}
\left.\frac{d}{d\gamma} \log p(y|\gamma)\right|_{\gamma=1} & \doteq (a-1)\log \tilde{\theta}_n+(b-1)\log(1-\tilde{\theta}_n)+O\left(\frac{1}{n}\right),
\end{align}
so that
\begin{align}
\left.\frac{d}{d\gamma} \log p(y|\gamma)\right|_{\gamma=1} & \doteq \log g(\tilde{\theta}_n)+O\left(\frac{1}{n}\right),
\end{align}
and this is a discrepancy that is asymptotically equivalent to the check suggested by \cite{evans+m06} (see also \cite{evans+j11b}).  
Again, it is easy to see following the argument of Section 3 that an arithmetic mixture involving the uniform will lead to the same result.  
So for appropriate
expansions of the $\mbox{Beta}(a,b)$ prior we can obtain checks asymptotically equivalent to both (\ref{emcheck}) and (\ref{klcheck}).
\end{exmp}

\begin{exmp}
{\it Normal location-scale model, hierarchically structured check}

\noindent
Consider $y=(y_1,\dots, y_n)\sim N(\mu 1_n,\sigma^2 I_n)$, where $\mu$ and $\sigma^2$ are both unknown, 
$1_n$ denotes an $n$-vector of ones and $I_n$ denotes the $n\times n$ identity matrix.  
The prior is $g(\mu,\sigma^2)=g(\mu|\sigma^2)g(\sigma^2)$, where $g(\sigma^2)=IG(a,b)$ ($IG(a,b)$ denotes the inverse gamma density with parameters $a$ and $b$) and $g(\mu|\sigma^2)=N\left(\mu_0,\frac{\sigma^2}{\lambda_0}\right)$.  Here $\mu_0$, $\lambda_0$, $a$ and $b$ are fixed hyperparameters.  This is an example of a hierarchically specified prior, and  $g(\mu,\sigma^2)$ is the conjugate normal inverse gamma prior for this problem. 

Consider checking the mean component of the prior, $g(\mu|\sigma^2)$.  We expand the prior $g(\mu|\sigma^2)$ to $g(\mu|\sigma^2,\lambda)=N\left(\mu_0,\frac{\sigma^2}{\lambda}\right)$, where now $\lambda$ is allowed to vary (i.e. it is no longer fixed at $\lambda_0$).  We have
\begin{align*}
  y|\sigma^2,\lambda & \sim N\left(\mu_0 1_n,\sigma^2\left(I_n+\frac{1}{\lambda}E_n\right)\right),
\end{align*}
where $E_n$ denotes an $n\times n$ matrix of ones.  Note that 
$$\left(I_n+\frac{1}{\lambda}E_n\right)^{-1}=I_n-\frac{1}{\lambda+n}E_n,$$
and 
\begin{align*}
  \log p(y|\sigma^2,\lambda) = &  -\frac{n}{2}\log 2\pi\sigma^2-\frac{1}{2}\log |I_n+\frac{1}{\lambda}E_n| \\
 & \;\; -\frac{1}{2\sigma^2}(y-\mu_01_n)^T\left(I_n+\lambda^{-1}E_n\right)^{-1}(y-\mu_0 1_n) \\
  = &-\frac{n}{2}\log2\pi\sigma^2-\frac{1}{2}\log |I_n+\frac{1}{\lambda}E_n | \\ 
  & \;\;-\frac{1}{2\sigma^2}\text{tr} \left((y-\mu_0 1_n)(y-\mu_0 1_n)^T\left(I_n-(\lambda+n)^{-1}E_n\right)\right),
\end{align*}
where $\text{tr}(\cdot)$ denotes the matrix trace.
This gives
\begin{align*}
\frac{d}{d\lambda} \log p(y|\sigma^2,\lambda) & = \frac{1}{2\lambda^2}\text{tr}\left((I_n+\lambda^{-1}E_n)^{-1}E_n\right)- \\
& \quad \quad \frac{1}{2\sigma^2}(y-\mu_0 1_n)^T((\lambda+n)^{-2}E_n)(y-\mu_0 1_n).
\end{align*}
Noting that 
\begin{align*}
  (y-\mu_01_n)^T((\lambda+n)^{-2}E_n)(y-\mu_0 1_n) & = \frac{n^2 (\bar{y}-\mu_0)^2}{(\lambda+n)^2},
\end{align*}
we obtain
\begin{align*}
  S^{(1)}(y) & \doteq (\bar{y}-\mu_0)^2,
\end{align*}
which is the Kullback-Leibler based check considered in \cite{nott+wee16}.  \cite{nott+wee16} also note
that the check is very similar to the one suggested in \cite{evans+m06}, p. 909.  

In this example we could have expanded the prior $g(\mu|\sigma^2)$ into the family $g(\mu|\sigma^2,\mu')=N(\mu',\sigma^2/\lambda_0)$, where $\mu'$ is not necessarily equal to $\mu_0$.  If we do this, we obtain
$S^{(1)}(y)\doteq (\bar{y}-\mu_0)$, and computation of a two-sided $p$-value gives that this is equivalent to a check using the statistic $(\bar{y}-\mu_0)^2$, so that the two
different ways of expanding the prior lead to the same result in this case.  An example where two different embedding families lead to useful and quite different answers is considered later.  
\end{exmp}

\section{More complex examples}

We now consider two complex examples which illustrate the main advantage of our method, which is that the choice
of a certain expansion can give conflict checks which are sensitive to conflicts of certain kinds.  
The first example considers checking
the appropriateness of a LASSO penalty in penalized regression using an exponential power prior, 
and shows that our method has improved performance
compared with an existing method in the literature which does not make use of the prior expansion family in the design
of the checking statistic.  
A second example is concerned with a problem in quantum tomography.  Here we consider
two different prior expansions, and show that for data simulated under the prior predictive for these priors
it is the check from the family used in the data simulation that is most effective for detecting conflict.  
These examples illustrate that score-based conflict checks based on appropriate prior expansions are helpful for
focusing checks on different aspects of the prior in complex situations. 

\subsection{Checking the appropriateness of a LASSO penalty}

A problem discussed in \cite{griffin+h17} is now
considered. The goal is to assess whether or not
a penalty term, used in a penalized regression to induce sparsity, is contradicted by
the data. Since the use of the penalty term they consider is equivalent to
employing a prior together with MAP estimation, checking the penalty term can
be addressed by prior-data conflict checking, and that is how we approach it here.

\begin{exmp}
{\it Many means problem with LASSO penalty}

\noindent
To start we restrict to the many means context with no
predictors, since the analysis is easier and the behavior reflects what
happens in the more general situation. Suppose $\bar{x}\sim N
(\mu,(\sigma^{2}/m)I_n)$ is observed with $\mu\in R^{n}$ and there is a belief
that $\mu$ is sparse, namely, many of the means satisfy $\mu_{i}=0$. It is
also assumed that $\sigma^{2}=1$ is known, as nothing material beyond
computational complexity is added to the analysis by placing a prior on this
quantity. For the prior on $\mu$, consider a product prior where each $\mu_{i}$
has density%
\begin{equation}
g(\nu\,|\,\tau,q)=\frac{q}{2\tau}\left(  \frac{\Gamma(3/q)}{\Gamma
(1/q)^3}\right)  ^{1/2}\exp\left\{  -\left(  \frac{\Gamma(3/q)}{\Gamma
(1/q)}\right)  ^{q/2}\left\vert \frac{\nu}{\tau}\right\vert ^{q}\right\},
\label{eq0}%
\end{equation}
for $\nu\in R^{1}$.  This is the exponential power family of priors that was considered
in \cite{griffin+h17}, and it can be shown that if $\mu_i$ has prior (\ref{eq0}), then 
$E(\mu_{i})=0$ and $\text{Var}(\mu_{i})=\tau^{2}$.  When $q=2$, 
the prior is normal, $N(0,\tau^2)$, and when $q=1$, the prior is a 
Laplace rescaled by $\tau/\sqrt{2}$.  
As $q\rightarrow0$ this family of priors induces greater sparsity.

A question of interest is whether or not the Laplace
prior obtained when $q=1$ conflicts with the data, as this corresponds to the popular LASSO penalty
\cite{tibshirani96}. \cite{griffin+h17} effectively compare the observed value of
the kurtosis statistic
\[
k(\bar{x})=\sum_{i=1}^{n}\bar{x}_{i}^{4}\bigg/ \left(  \sum_{i=1}^{x}\bar{x}_{i}%
^{2}\right)  ^{2},%
\]
with its prior distribution when $q=1$. Actually they use the prior
distribution of the kurtosis of a sample of $n$ from the prior itself as the
reference distribution for computational reasons, but we use the more appropriate prior distribution of
$k(\bar{x})$ for this comparison. If the observed $k(\bar{x})$ lies in the tails
of its prior predictive density, then this is an indication that the double exponential
prior is in conflict, and a modification of the prior is needed. The $p$-value value for the check is $2\min
\{P(k(\bar{X})<k(\bar{x})),P(k(\bar{X})>k(\bar{x}))\}$ where $P$ is the prior
predictive measure, and if we decide a 
conflict occurs when this $p$-value is less than $0.05$, then the left and right critical
values for testing $q=1$ are $(1.65,6.72)$ when $n=10$, and $(3.01,10.07)$
when $n=100$. So if $n=10$ and $k(\bar{x})<1.65$ or $k(\bar{x})>6.72$, then a prior-data conflict exists.

With $g(\mu|\bar{x},\tau,q)$ denoting the conditional posterior for $\mu$ given $\tau$ and $q$, the score function for
assessing sensitivity to $q$ is
\begin{equation}
S(\bar{x}\,|\,\tau)=\int g(\mu\,|\,\bar{x},\tau,1)\left.  \frac{d}%
{dq}\log g(\mu\,|\,\tau,q)\right\vert _{q=1}\,d\mu, \label{eq1}%
\end{equation}
namely, the posterior expectation of the derivative of $g(\cdot\,|\,\tau,q)$
with respect to $q$ evaluated at $q=1.$ A simple calculation leads to%

\begin{equation}
\left.  \frac{d}{dq}\log g(\mu\,|\,\tau,q)\right\vert _{q=1}=A(1)+B(1)\sum_{i=1}^{n}\left\vert \frac{\mu_{i}}{\tau}\right\vert +C(1)\sum_{i=1}^{n}\left\vert \frac{\mu_{i}}{\tau}\right\vert \log\left\vert \frac{\mu_{i}}{\tau}\right\vert, \label{eq2}
\end{equation}
where $A(1)=1+3(\psi(1)-\psi(3))/2$, $B(1)=-\Gamma(3)^{1/2} \left(  \log\Gamma(3)+\psi
(1)-3\psi(3)\right) /2$, and $C(1)=-\Gamma(3)^{1/2}$.  As previously, $\psi(x)$ denotes the digamma function.  Rather than computing the
expectation in (\ref{eq1}), this is approximated by $\hat{S}(\bar{x}%
\,|\,\tau)$ where the estimates\ $\mu_{i}=\bar{x}_{i}$ are substituted into
(\ref{eq2}). It is assumed hereafter that the elicited value of
$\tau$ is $\tau=1$.   In general $\tau$ is chosen such that the
effective support of the prior, which can be defined as a central interval
containing say $0.99$ of the prior probability when $q=1$, covers all the
$\mu_{i}$ values. 
Although the $\mu_i$ are not observed, in many applications we will have reliable prior knowledge of the plausible range of
location parameters which makes setting a prior scale parameter like $\tau$ relatively easy.  However, the choice
of the parameter $q$ controlling the heaviness of the prior tails is much more difficult, 
and so we focus on prior-data conflicts arising from the choice of $q$.
Figure \ref{fig1} is a plot of the null distribution of $\hat
{S}(\bar{x}\,|\,1)$ when $n=10$ and $m=20$, which leads to critical values
$(0.408,1.117).$ When $n=100$ and $m=20$, the critical values are given by
$(0.670,0.898)$.
\begin{figure}[ptb]
\begin{center}
\includegraphics[height=2.5in]{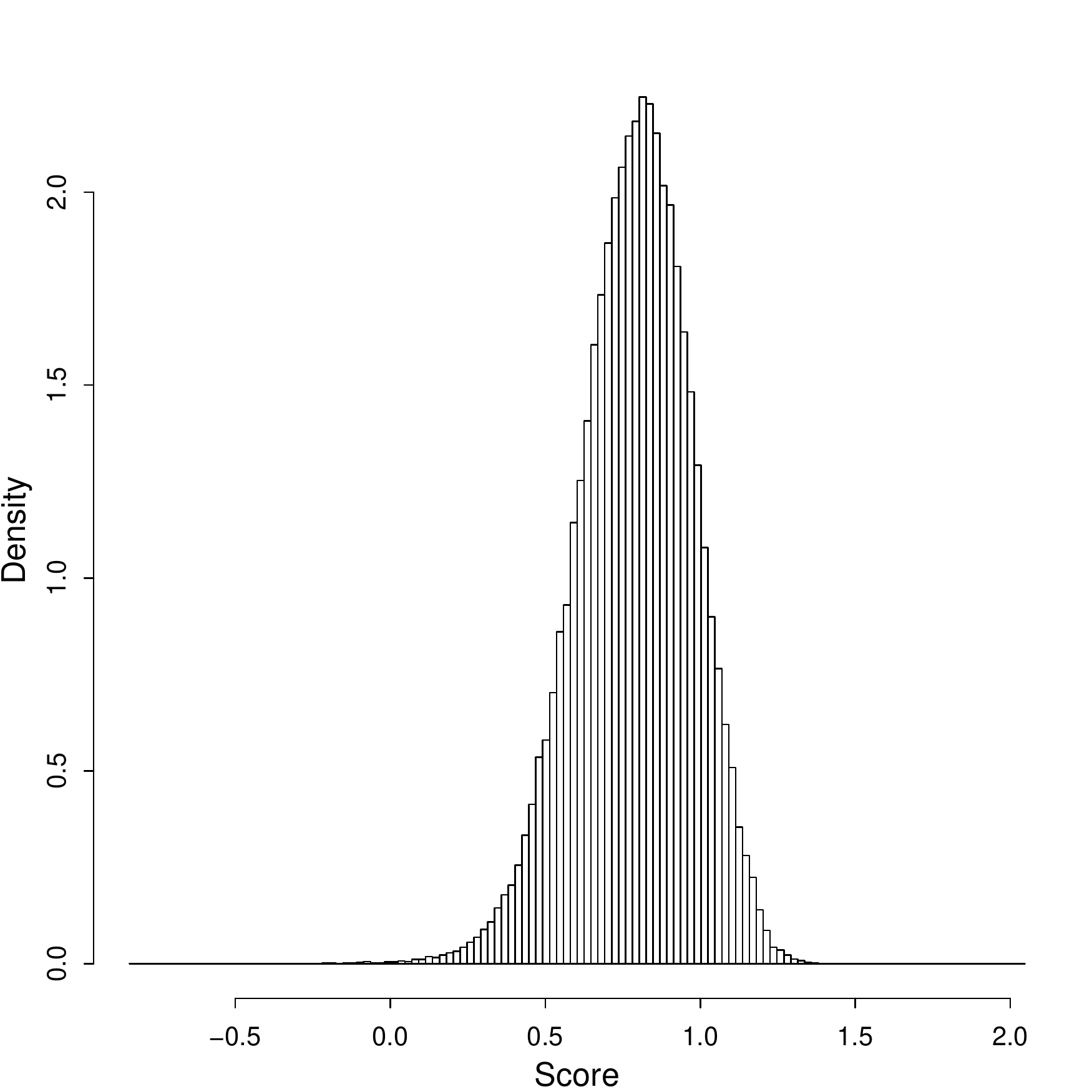}.
\caption{\label{fig1}Density histogram based on $10^{5}$ values generated from the prior
distribution of the approximate statistic based on $n=10,m=20,\tau=1$ .}
\end{center}
\end{figure}
To compare our approach with the method of \cite{griffin+h17}, the power of the tests was compared. 
By power we mean the following.  Consider a prior-predictive $p$-value such as (\ref{scorecheck}) and suppose a conflict
is declared if the $p$-value is less than $0.05$.  Then if data are simulated from the prior 
predictive density $p(y|\gamma)$, we can ask what is the probability that a conflict is detected?  Considering
this probability as a function of $\gamma$ gives a power function.  Figure 2 shows plots of the power functions of the kurtosis
and approximate score statistics in different situations where the expansion parameter $\gamma$ is $q$.  
It is seen that the approximate
score approach compares quite favorably with the method of \cite{griffin+h17}.
\end{exmp}
\begin{figure}[ptb]
\begin{center}
\begin{tabular}{c}
\includegraphics[height=2.5in]{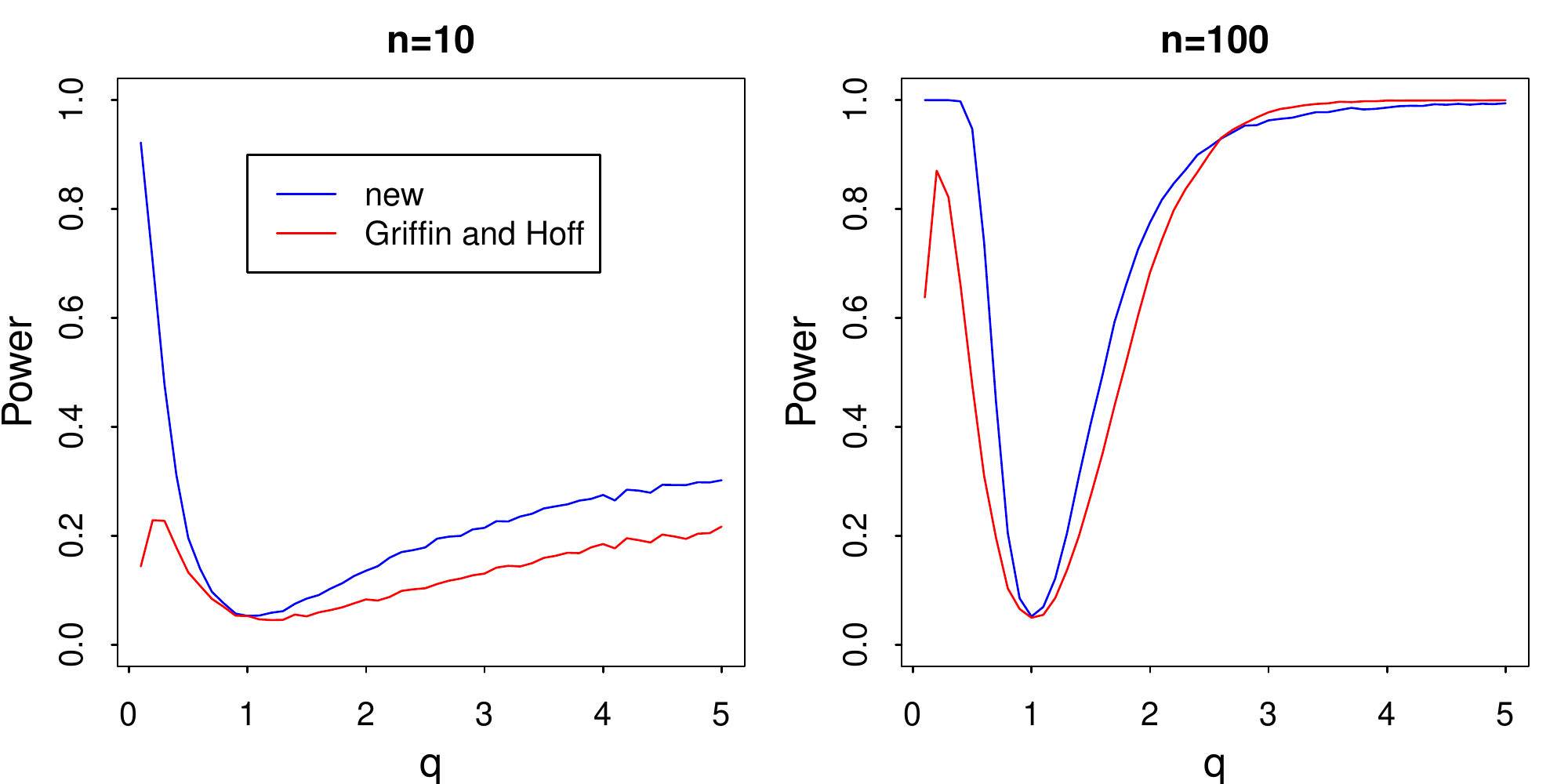} 
\end{tabular}
\caption{\label{fig2}Power functions when using the kurtosis statistic and
approximate score statistic for testing $q=1$ when $n=10$ (left) and $n=100$ (right) with $m=20$ and $\tau=1$.}
\end{center}
\end{figure}

\begin{figure}[ptb]
\begin{center}
\includegraphics[height=5in,width=5in]{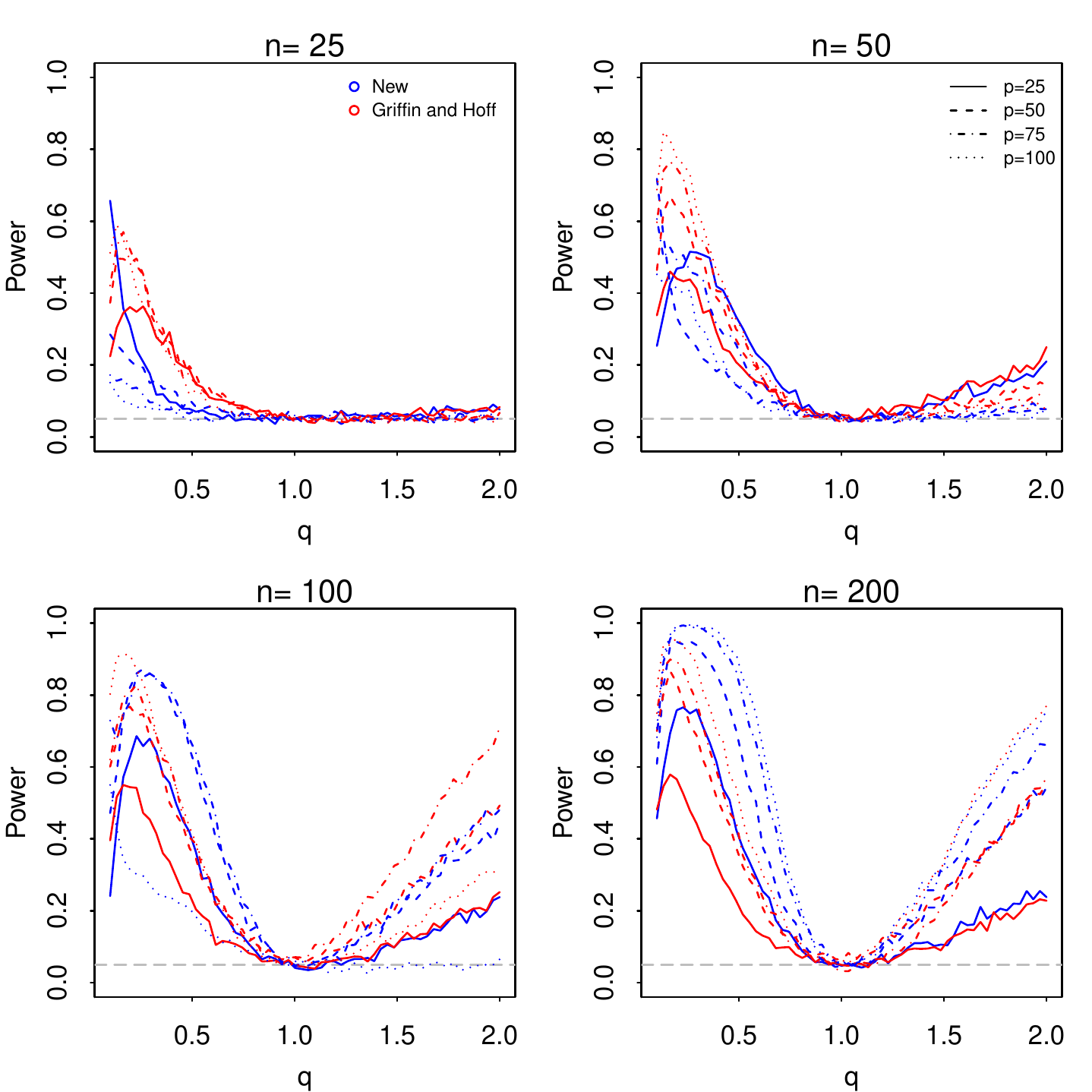}
\caption{\label{fig3}Simulation results to compare the powers of the kurtosis and
approximate score statistic.}%
\end{center}
\end{figure}

\begin{exmp}
{\it Regression with LASSO penalty}

\noindent
We now extend from the many means setting to a regression problem with data 
$y=X\beta+\sigma z$ where $X\in R^{n\times p}%
,\beta\in R^{p}$ is unknown, $z\sim N(0,I_n)$ and again $\sigma^{2}=1$ is
assumed$.$ Also for simplicity it is assumed that the $\beta_{i}$ can all be
treated equivalently so there is no intercept term which must be treated
differently. The prior on $\beta$ is taken to be a product prior with the same prior
(\ref{eq0}) placed on each $\beta_{i}.$ In practice, the columns of $X$ can be
standardized to each have sum 0 and unit length. With this assumption, the mean
of the $i$th coordinate of $y$ is $x_{i}^T\beta$, where $x_{i}^T$ 
is the $i$-th row of $X$, and note that $x_{i}\in\lbrack-1,1]^{p}$ because
of the standardization. As such, a bound on the means $x_{i}^T \beta$
that holds for all $i$ implies that $\tau$ can be chosen to guarantee that
the bounds on the means hold with high prior probability. Accordingly, our
concern for prior-data conflict can focus on $q$, and again the case $q=1$ is
considered.

For the kurtosis and approximate score, similar formulae are obtained as with
the many means case, and here the $\beta_{i}$ are estimated via least-squares.
When $k>n,$ so that $X$ is no longer of full rank, the Moore-Penrose estimates are
used as these minimize the length of the estimate vector and that seems
appropriate when considering sparsity. \cite{griffin+h17} used a ridge
estimator in the non-full rank case but this made little difference in the
results reported here.

A simulation study was conducted as in Section 2 of \cite{griffin+h17}.
Data were generated from the regression model with $\sigma^{2}=\tau^{2}=1$, for
$n=25,50,100$ and $200$ and $p=25,50,75$ and $100$. The entries of $X$ were
drawn from the standard normal distribution. For $10^{3}$ independent replicates of $X$ and
$\beta$ (drawn from the prior with $q=1$) the power was estimated for a grid
of values for $q$ from $0.1$ to $2$ in steps of $0.1.$ The cutoff level in the
test to determine the existence of a prior-data conflict was $0.05$. The
simulation results are given in Figure \ref{fig3}. It is seen that the approximate
score does quite well, and in certain cases, namely, when $p<n,$ can do better
than the test based on the kurtosis statistic.

The approximate score doesn't do as well as the kurtosis statistic when $p>n.$
This is felt in part to be due to the simulation performed. For when
generating $X$ via independent standard normals the matrix is of rank $n$ with
probability 1 when $p\geq n.$ So this situation is somewhat like having $n$
observations with $n$ independent variables and in such a case it is not
possible to criticize the model, as it will fit the data perfectly, let alone
the prior. In practice, if we wish to check both the prior and likelihood components of the model, and 
the coefficient vector is known to be sparse, then 
a preliminary screening of variables \cite{fan+lv18} can reduce $p$ to $p<n$ before penalized
regression is performed, and the score based check would seem to be preferable for checking the appropriateness
of the penalty in that case.  Such a screening procedure could also be implemented together with data splitting, where the
screening and analysis are done using disjoint subsets of the data.  
\end{exmp}

\newcommand{\half}{\frac{1}{2}}\newcommand{\thalf}{\tfrac{1}{2}}
\newcommand{\third}{\frac{1}{3}}\newcommand{\tthird}{\tfrac{1}{3}}
\newcommand{\ignore}[1]{\relax} 
\newcommand{\ONE}{\mathbf{1}} 
\newcommand{\MARKED}[1]{\textcolor{red}{#1}} 

\subsection{Checking a truncated Dirichlet prior in a constrained multinomial
  model for quantum state estimation}
The data acquired in measurements on quantum systems are fundamentally
probabilistic because -- as a matter of principle, not for lack of knowledge
-- one can only predict the probabilities for the various
outcomes but not which outcome will be observed for the next quantum system
to be measured.
Therefore, the interpretation of quantum-experimental data requires the use of
statistical tools, where the constraints that identify the
set of physically allowed probabilities must be enforced.
We shall illustrate prior checking in this context for a simple example, after
setting the stage by recalling some basic tenets of quantum theory.

In the formalism of quantum theory, the Hilbert space operators of a
$D$-dimensional quantum system can be represented by $D\times D$ matrices;
for simplicity, we shall not distinguish between the operators and the
matrices that represent them.
There is, in particular, the statistical operator $\rho$ that describes the
state of the quantum system, which is positive-semidefinite and has unit
trace.
A measurement with $K$ outcomes is specified by $K$ positive-semidefinite
probability operators (also commonly referred to as POVMs) $\Pi_1$, $\Pi_2$, \dots, $\Pi_K$, 
one for each outcome.
When such a measurement is performed on independent and identically prepared
systems, we get a click of one of the detectors for each of the measured
systems, and the probability that the $k$th detector will click for the next
quantum system is $\theta_k=\tr(\rho\Pi_k)$ (``Born's rule'').
The unit sum of the probabilities, $\sum_{k=1}^K\theta_k=1$, is ensured for any
$\rho$ by the unit sum of the $\Pi_k$s, $\sum_{k=1}^K\Pi_k=I_D$, where
as previously $I_D$ denotes the $D\times D$ identity matrix.

The $K$-tuplets of probabilities, $\theta=(\theta_1,\theta_2,\dots,\theta_K)$,
constitute a convex set in the $(K-1)$-simplex but usually they do not exhaust
this simplex (for an example, see Figure~\ref{fig:trine-simplex} below).
The permissible probabilities are those consistent with $\rho\geq0$, and the
actual constraints obeyed by the $\theta_k$s result from the properties of the
$\Pi_k$s.
We denote the convex set of permissible $\theta$s by $\Theta$.

After measuring $N$ identically prepared copies of the quantum system, thereby
counting $n_k$ clicks of the $k$th detector, we have the data
$y=(n_1,n_2,\dots,n_K)$ with $\sum_{k=1}^Kn_k=N$.
The problem of inferring the statistical operator $\rho$ from the data $y$ is the
central theme of quantum state estimation \cite{paris+r04,teo15}, 
and Bayesian methods are well-suited for this task \cite{shang13,xikun16}.
Enforcing $\rho\geq0$, or the corresponding implied constraints on the
probabilities $\theta$, is crucial and often challenging.
We shall consider a rather simple example below, with ${D=2}$ and ${K=3}$.

The data $y$ are modeled as multinomial with parameter
$\theta=(\theta_1,\theta_2,\dots,\theta_K)\in\Theta$. 
We recall that a random vector $\Delta=(\Delta_1,\Delta_2,\dots, \Delta_K)$
follows a Dirichlet distribution with parameter $(\alpha_1,\alpha_2,\dots,
\alpha_K)$, denoted by $\text{Dir}(\alpha_1,\dots, \alpha_K)$, if it has density 
$$g(\Delta)\propto \prod_{k=1}^K \Delta_k^{\alpha_k-1}$$
on the $(K-1)$-dimensional region
$\{\Delta:  \Delta_k\geq 0 \mbox{\ for\ } k=1,\dots, K-1
\mbox{\ with\ }\sum_{k=1}^{K-1}\Delta_k\leq 1\}$
where $\Delta_K=1-\sum_{k=1}^{K-1} \Delta_k$.
Consider a prior for $\theta$ that is proportional to a  
Dirichlet prior $\text{Dir}(\alpha_0 q_1,\dots, \alpha_0 q_K)$ on $\Theta$
where $q=(q_1,\dots, q_K)$ is a location parameter,
$\sum_{k=1}^Kq_k=1$, and $\alpha_0>0$ is an overall precision parameter.
That is, the prior is
\begin{equation*}
  g(\theta|\alpha_0,q)  \propto \prod_{k=1}^K \theta_k^{\alpha_0q_k-1}
    \quad\mbox{for}\quad\theta\in \Theta.
\end{equation*}
The hyperparameters can be set by eliciting a point
estimate of $\theta$ as the value for $q$, and calibrating the precision parameter according
to the desired prior uncertainty about $\theta$.

\subsubsection{Two prior expansions}
We first consider two different families of expansions of the above prior,
which do not attempt to check violations of the physical constraint on
$\theta$.
Later we consider an expansion suitable for checking the physical constraint
in a simple situation ($D=2$ and $K=3$),
and where the formula (\ref{scorestat2}) does not hold without some modification. To minimize notation, in all the prior families considered
below the expansion parameter in the prior is denoted by $\gamma$, although it should be noted that in different families the interpretation
of this parameter differs. 

In our first prior expansion, similar to our earlier binomial example, we
consider a geometric mixture of the original prior with the Jeffreys' prior,
which is the Dirichlet prior $\text{Dir}(\half,\half,\dots,\half)$ constrained to
$\theta\in \Theta$.
Mixing with the Jeffreys' prior thickens the tails of the original prior, and
as such this family may be helpful for constructing an overall test of
conflict.
This idea leads to a family that is still constrained Dirichlet, proportional
to $\text{Dir}(\delta_1,\dots, \delta_K)$ with
$\delta_k=(1-\gamma) \alpha_0 q_k+\half\gamma$.  
The corresponding density function is denoted by
\begin{equation*}
  g^{(1)}(\theta|\alpha_0,q,\gamma)
  \propto \prod_{k=1}^K \theta_k^{\delta_k-1}
  \quad\mbox{for}\quad\theta\in \Theta.
\end{equation*}
The original prior is obtained when $\gamma=0$.

Our second prior family also derives from considering a constrained
Dirichlet density, $\text{Dir}(\delta'_1,\dots, \delta'_K)$, with 
$\delta_k'=\alpha_0q'_k$, $q_1'=q_1+\gamma$ and $q_k'=q_k-\gamma/(K-1)$
for $k=2,\dots, K$.
We see that $\sum_{k=1}^K q_k'=1$ and $\sum_{k=1}^K \delta_k'=\alpha_0$, so that
the overall precision parameter is kept fixed while changing the location
parameters by increasing $q_1$ by $\gamma$, with the other $q_k'$ adjusted to
maintain the unit-sum constraint.  
This family of priors is constructed to focus particularly on conflicts
involving the first component $\theta_1$ of $\theta$, and we obtain the
original prior at $\gamma=0$.
For this family we write
\begin{equation*}
  g^{(2)}(\theta|\alpha_0,q,\gamma)
   \propto \prod_{k=1}^K \theta_k^{\delta_k'-1}
   \quad\mbox{for}\quad\theta\in \Theta.
\end{equation*}

\subsubsection{Score statistics for the checks}
Consider first the family $g^{(1)}$.
Apart from terms not depending on $\theta$, we have
\begin{align*} 
  \frac{d}{d\gamma} \log g^{(1)}(\theta|\alpha_0,q,\gamma)
  & = \frac{d}{d\gamma} \sum_{k=1}^K (\delta_k-1)\log \theta_k, \\
  & = \sum_{k=1}^K (\alpha_0q_k-\thalf)\log \theta_k
\end{align*}
and, upon using (\ref{scorestat2}),
\begin{equation*}
 S(y)  \doteq \sum_{k=1}^K(\alpha_0 q_k-\thalf)E(\log \theta_k|y).
\end{equation*}
On the other hand, for the family $g^{(2)}$, we obtain that, apart from terms
not depending on~$\theta$, 
\begin{align*}
  \frac{d}{d\gamma} \log g^{(2)}(\theta|\alpha_0,q,\gamma)
  & =\frac{d}{d\gamma} \left( \sum_{k=1}^K (\delta_k'-1)\log \theta_k\right) \\
 & = \alpha_0 \log \theta_1-\alpha_0 \sum_{k=2}^K \frac{\log \theta_k}{K-1},
\end{align*}
and using (\ref{scorestat2}) again, 
\begin{equation*}
  S(y)  \doteq  E(\log \theta_1|y)-\sum_{k=2}^K \frac{E(\log \theta_k|y)}{K-1}.
\end{equation*}
We see that in the case of this prior family the components are no longer
treated symmetrically in the check, with conflicts involving $\theta_1$ being
the focus.

\begin{figure}[ptb]
\begin{center}
  \includegraphics[width=\textwidth]{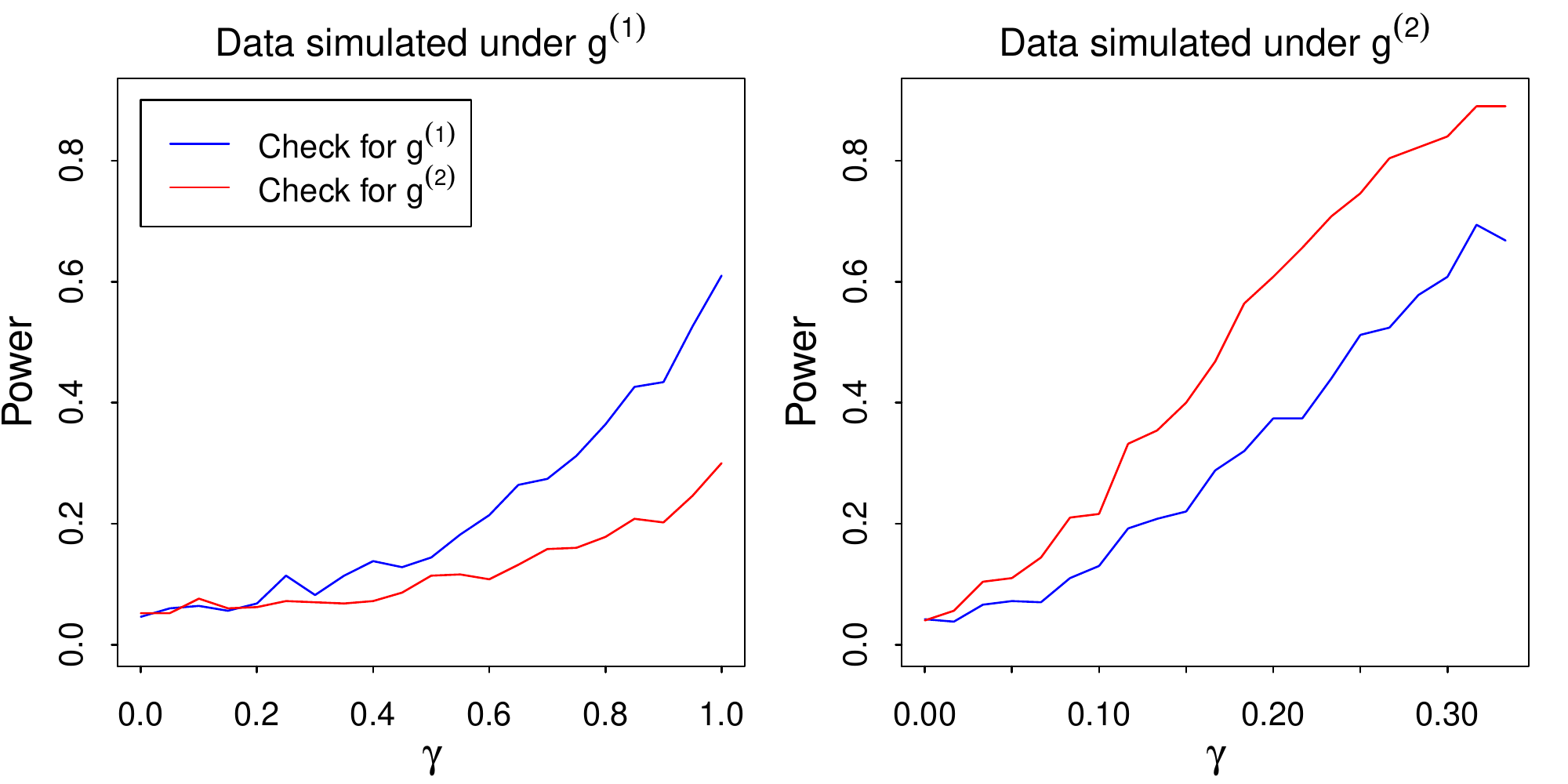}
  \caption{\label{fig4}%
    Power of checks based on families $g^{(1)}$ and $g^{(2)}$ when data are
    simulated under the prior predictive for $g^{(1)}$ (left) and under the
    prior predictive for $g^{(2)}$ (right).}
\end{center}
\end{figure}

\subsubsection{Power comparison for the checks}
We consider the case of ${K=3}$ with ${q=(\third,\third,\third)}$, and
$\alpha_0=30$.
For the family $g^{(1)}$, $\gamma=0$ corresponds to a $\text{Dir}(10,10,10)$ prior, and
$\gamma=1$ to a $\text{Dir}(\half,\half,\half)$ prior.
For the $g^{(2)}$ expansion, $\gamma=0$ is a $\text{Dir}(10,10,10)$ prior, and
$\gamma=\third$ is a $\text{Dir}(20,5,5)$ prior.
We examine the power of the checks in two cases, where a $p$-value smaller
than $0.05$ is considered to be a conflict.
In the first case, we consider simulating data under the prior predictive for
the $g^{(1)}$ expansion, for values of $\gamma=i/20$, $i=0,\dots, 20$.
Figure~\ref{fig4} (left) shows how the power of the two checks varies with
$\gamma$, and we note that the check based on $g^{(1)}$ is more powerful, when
the prior predictive for $g^{(1)}$ is used for simulating the data.
The power is approximated at each value of $\gamma$ based on $500$ simulations.  
Next, we consider simulating data under the prior predictive for $g^{(2)}$,
for values of $\gamma=i/60$, $i=0,\dots, 20$.
Again $500$ simulations are performed at each $\gamma$ to compare power for
the two checks, and Figure \ref{fig4} (right) shows again that it is the check
corresponding to the family used to generate the data that is more powerful.
The different families are powerful against different kinds of conflict with
the original prior, and the expansion of the prior used can be constructed
with this in mind.

\subsubsection{A simple quantum measurement scenario}
The simplest genuine quantum system is that of a binary alternative
(${D=2}$), a \emph{qubit}.
Here, we have ${2\times2}$ matrices for all operators, and the usual
parameterization of the statistical operator is
\begin{equation*}
  \rho=\half{\left(\begin{array}{cc}
       1+s_3 & s_1-is_2 \\ s_1+is_2 & 1-s_3
      \end{array}\right)}
\quad\mbox{with}\quad s_1^2+s_2^2+s_3^2\leq1.
\end{equation*}
If we regard, as we shall, the three real parameters ${s_1,s_2,s_3}$ as
Cartesian coordinates of a point, then there is a one-to-one correspondence
between the quantum states of a qubit and the three-dimensional unit ball. 

Two-outcome measurements on qubits realize the situation of coin tossing and
do not exhibit features particular to quantum physics.
We shall, therefore, consider 3-outcome measurements (${K=3}$), for which we
choose the $\Pi_k$'s in accordance with
\begin{equation*}
  \Pi_k=w_k{\left(\begin{array}{cc}
       1 & e^{-i\phi_k} \\ e^{i\phi_k} & 1
            \end{array}\right)}
\quad\mbox{with}\quad w_k>0 \quad\mbox{for}\quad k=1,2,3.              
\end{equation*}
The corresponding probabilities
\begin{equation}\label{eq:qb-0}
  \theta_k=w_k(1+s_1\cos\phi_k+s_2\sin\phi_k)
\end{equation}
do not involve $s_3$, so that no information about this state parameter is
gained from such a measurement, and the three $\theta_k$s are restricted by
${s_1^2+s_2^2\leq1}$.
The relevant parameter space is now the unit disk in the $s_1,s_2$ plane, the
intersection of this plane with the unit ball.

\begin{figure}
  \centerline{\includegraphics[width=80mm]{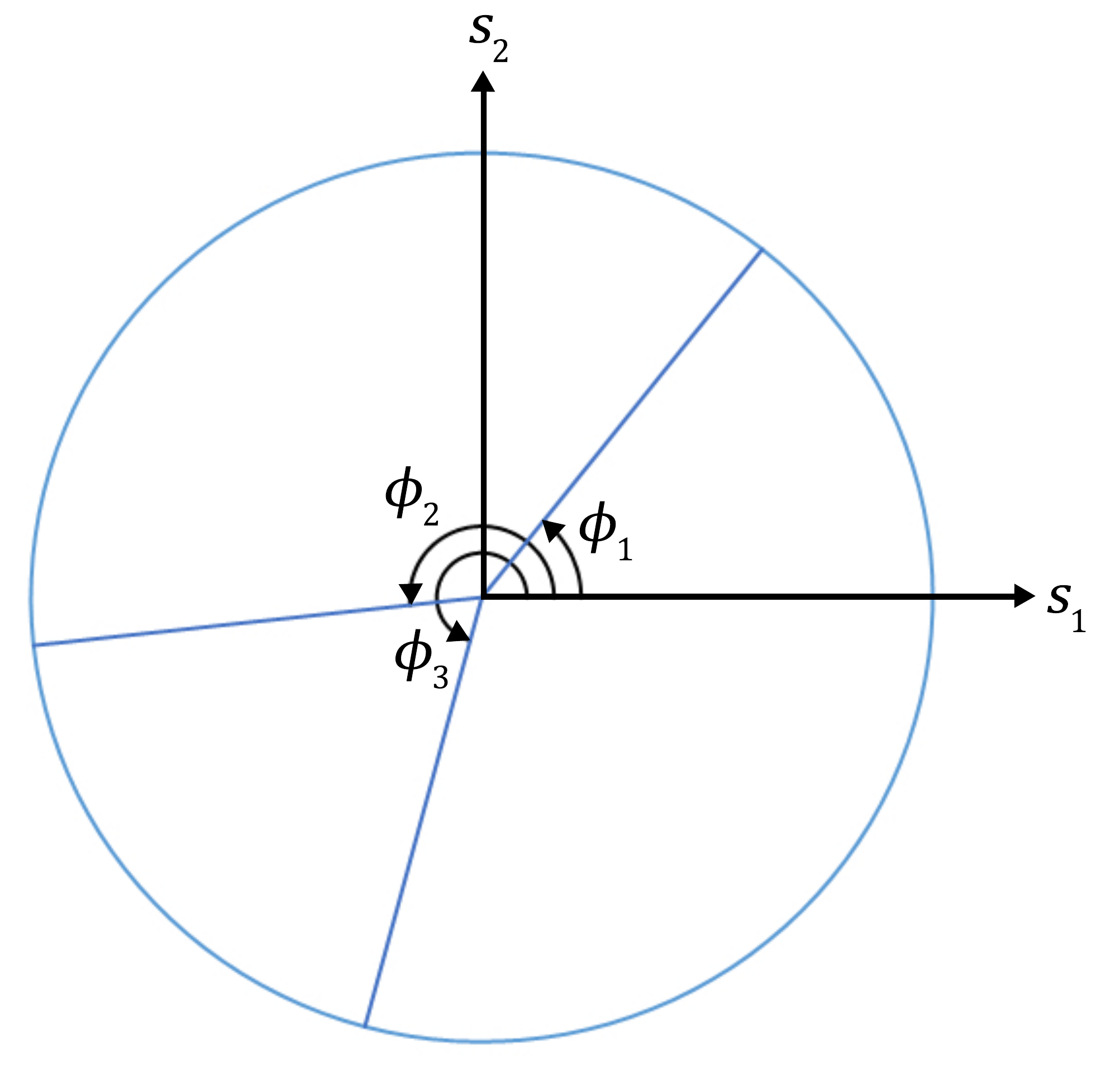}}
\caption{\label{fig:sliced-pie}%
  The angles $\phi_1$, $\phi_2$, $\phi_3$ in the probabilities (\ref{eq:qb-0})
  slice the circular unit-disk pie into three pieces, each slice smaller than
  half of the pie.} 
\end{figure}

The angles $\phi_1,\phi_2,\phi_3$ divide the unit disk into three pie slices;
see Figure~\ref{fig:sliced-pie}.
The condition
\begin{equation*}
  \Pi_1+\Pi_2+\Pi_3
  ={\left(\begin{array}{cc}1&0\\0&1\end{array}\right)}
\end{equation*}
or, equivalently,
\begin{equation*}
  \theta_1+\theta_2+\theta_3=1\quad\mbox{for all $s_1,s_2$}
\end{equation*}
determines the weights $w_k$ and they will be positive if each slice is
less than half of the pie; we take this for granted.

The symmetric case of three pie slices of equal size is that of the so-called
trine measurement.
Upon choosing ${\phi_1=0}$ by convention, we then have
${\phi_2=\frac{2}{3}\pi}$ and ${\phi_3=\frac{4}{3}\pi}$, and the probabilities
of the trine measurement are
\begin{equation}\label{eq:qb-trine}
  \theta_1=\third(1+s_1),\qquad
  \left.\begin{array}{l}\theta_2\\\theta_3\end{array}\right\}
  =\frac{1}{6}(2-s_1\pm\sqrt{3}s_2),
\end{equation}
which are constrained by
\begin{equation}\label{eq:qb-1}
  s_1^2+s_2^2=(3\theta_1-1)^2+3(\theta_2-\theta_3)^2\leq1.
\end{equation}
In view of ${\theta_1+\theta_2+\theta_3=1}$, this can be equivalently, and
more symmetrically, written as ${\theta_1^2+\theta_2^2+\theta_3^2\leq\half}$.

For ${\phi_1=0}$, ${\phi_2=\pi-\varphi}$, ${\phi_3=\pi+\varphi}$ we get a
symmetrically distorted trine, for which the probabilities are
\begin{equation}\label{eq:qb-2}
  \theta_1=\half(\sin\gamma)^2(1+s_1),\qquad
  \left.\begin{array}{l}\theta_2\\\theta_3\end{array}\right\}
  =\frac{1}{4}\bigl[1+(\cos\gamma)^2-s_1(\sin\gamma)^2\pm2s_2\cos\gamma\bigr]
\end{equation}
with ${\cos\gamma=\tan(\half\varphi)}$, and the analog of (\ref{eq:qb-1})
reads
\begin{equation}\label{eq:qb-3}
  {\left(\frac{2\theta_1}{(\sin\gamma)^2}-1\right)}^2
  +{\left(\frac{\theta_2-\theta_3}{\cos\gamma}\right)}^2\leq1
  \quad\mbox{for}\quad (\theta_1,\theta_2,\theta_3)\in\Theta_{\gamma},
\end{equation}
where we note that the set of permissible $\theta$s depends on the value of
$\gamma$. 
[Note: Later this distortion parameter $\gamma$ will play the role of the
generic expansion parameter $\gamma$.]
We recover the ideal trine for ${\varphi=\third\pi}$
and ${(\cos\gamma)^2=\third}$, and the limiting
cases of ${\varphi=\half\pi}$ and ${\varphi=0}$ yield degenerate 2-outcome
measurements of no further interest.
In the recent experiment by  \cite{len+ek17} different symmetrically
distorted trines were realized (for measuring the polarization qubit of a
photon), among them ${(\cos\gamma)^2=0.1327}$
for which ${y=(n_1,n_2,n_3)=(180,31,30)}$ were the counts of detection events.
While the actual counts in the experiments were about ten times as many,
namely $(1802,315,303)$ as communicated by author Y.~L.~Len, we are using
these smaller numbers here because prior-data conflicts are less of an issue
in data-dominated situations. However, even if posterior inferences are 
insensitive to a prior-data conflict in large data settings, it is 
still of interest to detect the conflict, since this indicates a lack 
of scientific understanding in setting up the model.

\begin{figure}
  \centerline{\includegraphics[width=100mm]{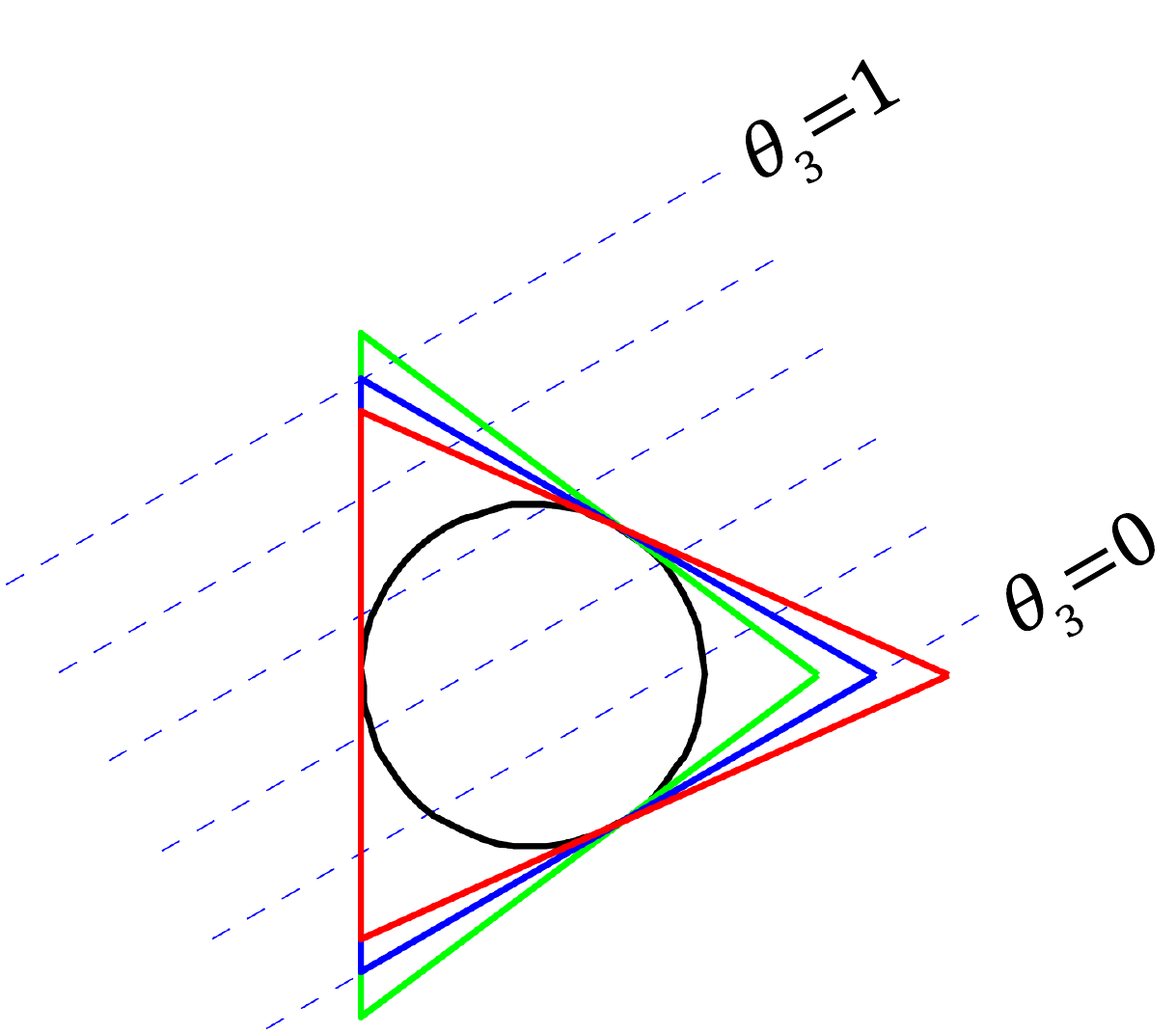}}
\caption{\label{fig:trine-simplex}%
  For the probabilities $(\theta_1,\theta_2,\theta_3)$ of the symmetrically
  distorted trine measurement in (\ref{eq:qb-2}), the physically allowed
  values correspond to the points on the unit disk (on and inside the black
  unit circle, the blue circle in Fig.~\ref{fig:sliced-pie}) while the probability
  2-simplex is a triangle whose sides touch the unit circle.
  The probabilities associated with top left and bottom left vertices of the
  triangle are $(\theta_1,\theta_2,\theta_3)=(0,0,1)$ and $(0,1,0)$,
  respectively, and the vertex on the right has $(1,0,0)$.
  The blue equilateral triangle is for the ideal-trine probabilities in
  (\ref{eq:qb-trine}) when $(\cos\gamma)^2=\third$, the green triangle is for
  $(\cos\gamma)^2=\third-\frac{1}{12}$, and the red triangle is for
  $(\cos\gamma)^2=\third+\frac{1}{12}$.
  The dashed blue lines show where $\theta_3=0,0.2,0.4,0.6,0.8,1$ for the blue triangle.}
\end{figure}

The probability space for the symmetrically distorted trine has a simple
geometry, illustrated in Figure~\ref{fig:trine-simplex}.
The unit disk in the $s_1,s_2$ plane accounts for all $s_1,s_2$ pairs for
which the probabilities in (\ref{eq:qb-2}) obey the constraint in
(\ref{eq:qb-3}); that is: the unit disk represents the set $\Theta$ of permissible
probabilities. 
The $s_1,s_2$ pairs for which one of the probabilities in (\ref{eq:qb-2}) has
a chosen value, mark a line in the $s_1,s_2$ plane, and different fixed values
for the same $\theta_k$ yield a set of parallel lines.
The particular three lines with ${\theta_1=0}$ or ${\theta_2=0}$ or
${\theta_3=0}$ are tangential to the unit circle and intersect where either
${(\theta_1,\theta_2,\theta_3)=(1,0,0)}$ or $(0,1,0)$ or $(0,0,1)$;
the triangle thus defined is the 2-simplex for the probabilities in
(\ref{eq:qb-2}).
For the ideal trine, it is an equilateral triangle; for the symmetrically
distorted trine, we have an isosceles triangle with vertices at
$(s_1,s_2)=(2(\sin\gamma)^{-2}-1,0)$ and $(-1,\mp(\cos\gamma)^{-1})$.
For the general case of (\ref{eq:qb-0}), there is an analogous construction
with a triangle with no particular symmetry for the 2-simplex. 
Among all these triangles, the equilateral triangle for the ideal trine has
the smallest area.

\subsubsection{A prior family for checking the physical constraints}
We now consider the symmetrically distorted trine with its $\gamma$-dependent
set of permissible $\theta$s, in accordance with (\ref{eq:qb-3}).
If it is suspected that the trine measurement set-up was not properly
balanced, this gives a natural family of priors for performing our check.

In such a situation where the support for the prior changes with $\gamma$,
(\ref{scorestat2}) 
can no longer be used to compute $S(y)$ without some modification.
Instead of using (\ref{scorestat2}), we expand (\ref{scorestat}) as 
\begin{align*}
  S(y) &=\left.\frac{d}{d\gamma}\log p(y|\gamma)\right|_{\gamma=\gamma_0}
         =\frac{\left.\frac{d}{d\gamma}p(y|\gamma)\right|_{\gamma=\gamma_0}}
         {p(y|\gamma_0)}  \,,
\end{align*}
with 
\begin{align*}
p(y|\gamma) &= \int\limits_{\Theta_{\gamma}}
               p(y|\theta)g(\theta|\gamma)\;d\theta
              = \int\limits_{\Theta_{\gamma}}  \binom{N}{n_1,n_2,n_3}
              \theta_1^{n_1}\theta_2^{n_2}\theta_3^{n_3}g(\theta|\gamma)
              \;d\theta  \,.
\end{align*}
The changing support makes $\frac{d}{d\gamma}p(y|\gamma)$ inconvenient to
evaluate numerically.
To deal with this, we switch from integrating over $\theta$ to integrating
over $s_1$ and $s_2$ and use polar coordinates in the $s_1,s_2$ plane,
\begin{equation*}
  s_1=r\cos\phi,\quad s_2=r\sin\phi
\end{equation*}
with ${0\leq r\leq1}$ and ${0\leq\phi\leq2\pi}$, to enforce the constraints
(\ref{eq:qb-2}) for any value of $\gamma$.
With this, we now have a situation where the sampling model itself changes
with $\gamma$: 
\begin{equation*}
  p(y|r,\phi,\gamma)=\binom{N}{n_1,n_2,n_3}
    \theta_1^{n_1}\theta_2^{n_2}\theta_3^{n_3}
    \Biggr|_{\mbox{\footnotesize\begin{tabular}[t]{@{}l@{}}%
                   $\theta_k$ from (\ref{eq:qb-2})\\
                   with $s_1+is_2=re^{i\phi}$\end{tabular}}}.
\end{equation*}

\begin{figure}
  \centerline{\includegraphics[scale=1.0,viewport=105 240 490 545,clip=,width=100mm]%
                       {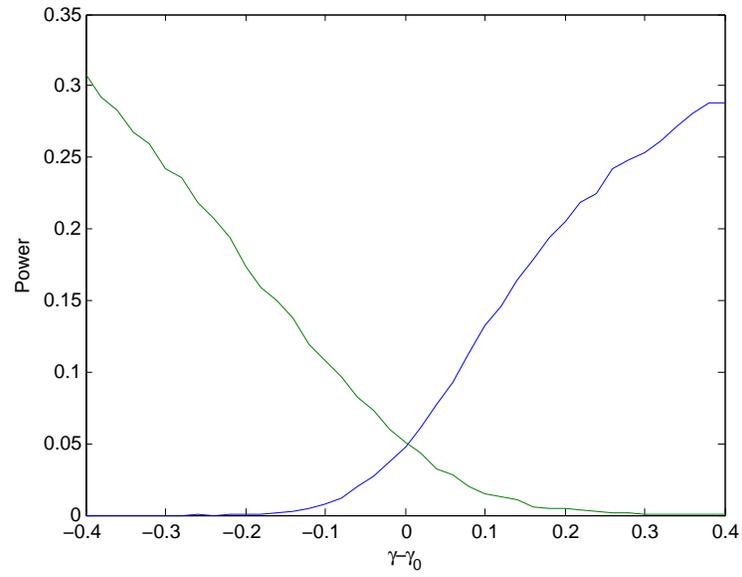}}
\caption{\label{fig:power-flat}%
  Probability of detecting a conflict at a $p$-value threshold $0.05$
  for data simulated under the prior predictive for different prior
  hyperparameter $\gamma$, for the case where $\cos^2(\gamma_0)=\third$ and
  $N=50$.
  The blue and green lines show power for increasing and decreasing $\gamma$,
  respectively.} 
\end{figure}

\paragraph{Dirichlet prior over the physical space}
Consider the case of a Dirichlet prior over the physical space, i.e.,
$g(\theta|\gamma)\propto %
\theta_1^{\alpha_1-1}\theta_2^{\alpha_2-1}\theta_3^{\alpha_3-1}I_\gamma(\theta)$,
where $I_\gamma(\theta)$ is the indicator function,
\begin{equation*}
      I_\gamma(\theta)= 
\begin{cases}
    1 & \text{if $\theta$ is permissible: $\theta\in\Theta_{\gamma}$,}\\
    0 & \text{otherwise: $\theta\not\in\Theta_{\gamma}$.}
\end{cases}
\end{equation*}
Under the $r,\phi$ parameterization, we then have 
\begin{equation*}
  S(y)=\frac{d}{d\gamma}\log\frac{H_1(\gamma)}{H_2(\gamma)}
  \Biggr|_{\gamma=\gamma_0} 
\end{equation*} 
where
\begin{equation*}
  H_1(\gamma)
  = \int\theta_1^{n_1+\alpha-1}\theta_2^{n_2+\alpha-1}\theta_3^{n_3+\alpha-1}
    \Biggr|_{\mbox{\footnotesize\begin{tabular}[t]{@{}l@{}}%
                   $\theta_k$ from (\ref{eq:qb-2})\\
                   with $s_1+is_2=re^{i\phi}$\end{tabular}}}
             r dr\, d\phi
\end{equation*}
and
\begin{equation*}
H_2(\gamma)= \int\theta_1^{\alpha-1}\theta_2^{\alpha-1}\theta_3^{\alpha-1}
    \Biggr|_{\mbox{\footnotesize\begin{tabular}[t]{@{}l@{}}%
                   $\theta_k$ from (\ref{eq:qb-2})\\
                   with $s_1+is_2=re^{i\phi}$\end{tabular}}}
 r dr \,d\phi.
\end{equation*} 
The two-dimensional integrals above can be performed numerically.
Figure~\ref{fig:power-flat} shows the power of the conflict score test when
the underlying prior has a $\gamma$ value deviating from $\gamma_0$ for the
case of a flat prior, where $\alpha=\{1,1,1\}$.
Here, the two curves are obtained by flipping the sign of the score
function.
We see that they behave as we expect them to, one being sensitive
to conflicts caused by $\gamma$ values too large, the other by values too small.

\paragraph{An example from a quantum experiment}
As mentioned above, symmetrically distorted trines were realized in the
experiment recently conducted by \cite{len+ek17}.
We now consider the data $y=(180,31,30)$ for the distorted trine with
$(\cos\gamma)^2=0.1327$.
Suppose a prior which is flat over the symmetric trine is chosen
($(\cos\gamma)^2=\third$).
Then the score-based conflict check gives a $p$-value of $0.00004$,
indicating a conflict.
If instead the correct $\gamma$ is chosen, the same test yields a $p$-value of
$0.56$.

\section{Discussion}
We have considered a new approach to constructing prior-data conflict checks based on embedding the prior used for the analysis into a larger family and then considering a marginal likelihood score statistic for the expansion parameter.  The main advantage of this technique is that through the choice of the prior expansion we can construct checks which are sensitive to different aspects of the prior.    

There are a number of ways in which our work could be extended.  In Section 4, we considered checking for the appropriateness of the LASSO penalty in penalized regression, but it would be interesting also to check other commonly used sparse signal shrinkage priors.  For example, the generalized Beta mixture of Gaussians family of \cite{armagan+cd11} would provide
a suitable prior expansion for checking the horseshoe prior \cite{carvalho+ps09,carvalho+ps10} in our framework.  It would also be interesting to use the score-based approach for checking priors on hyperparameters in nonparametric models like Gaussian processes.  In the nonparametric setting, priors can be crucial for limiting flexibility and avoiding overfitting, but it can also be difficult to understand the predictive implications of an informative prior which makes checking the prior important.

For complex hierarchical priors it is challenging to implement conflict checking methods computationally in an automatic way.  A promising recent
development in this direction is the work of \cite{seth+mw19}, and building on earlier work of \cite{yuan+j12}.  \cite{seth+mw19}  consider a comparison of a single draw from the posterior with the prior distribution and exploiting any exchangeable structure in the prior in the comparison, and explain why their approach gives well-calibrated $p$-values.  
We believe it is possible to combine our score-based checks with this idea, and implementation would be relatively easy
to do with standard statistical software.  However, the method of \cite{seth+mw19} is a randomized method, and there may
be a statistical price to be paid for the convenient implementation it provides.  Investigation of this is left to future work.


\begin{thebibliography}{44}
\newcommand{\enquote}[1]{``#1''}
\expandafter\ifx\csname natexlab\endcsname\relax\def\natexlab#1{#1}\fi
\expandafter\ifx\csname url\endcsname\relax
  \def\url#1{{\tt #1}}\fi
\expandafter\ifx\csname urlprefix\endcsname\relax\def\urlprefix{URL }\fi
\ifx\endbibitem\undefined \let\endbibitem\relax\fi

\bibitem[{Al~Labadi and Evans(2017)}]{allabadi+e17}
Al~Labadi, L. and Evans, M. (2017).
\newblock \enquote{Optimal Robustness Results for Relative Belief Inferences
  and the Relationship to Prior-Data Conflict.}
\newblock {\em Bayesian Analysis\/}, 12(3): 705--728.
\endbibitem

\bibitem[{Armagan et~al.(2011)Armagan, Clyde, and Dunson}]{armagan+cd11}
Armagan, A., Clyde, M., and Dunson, D.~B. (2011).
\newblock \enquote{Generalized Beta Mixtures of {G}aussians.}
\newblock In Shawe-Taylor, J., Zemel, R.~S., Bartlett, P.~L., Pereira, F., and
  Weinberger, K.~Q. (eds.), {\em Advances in Neural Information Processing
  Systems 24\/}, 523--531.
\endbibitem

\bibitem[{Baskurt and Evans(2013)}]{baskurt+e13}
Baskurt, Z. and Evans, M. (2013).
\newblock \enquote{Hypothesis Assessment and Inequalities for {B}ayes Factors
  and Relative Belief Ratios.}
\newblock {\em Bayesian Analysis\/}, 8(3): 569--590.
\endbibitem

\bibitem[{Bayarri and Berger(2000)}]{bayarri+b00}
Bayarri, M.~J. and Berger, J.~O. (2000).
\newblock \enquote{P Values for Composite Null Models (with discussion).}
\newblock {\em Journal of the American Statistical Association\/}, 95: pp.
  1127--1142.
\endbibitem

\bibitem[{Bayarri and Castellanos(2007)}]{bayarri+c07}
Bayarri, M.~J. and Castellanos, M.~E. (2007).
\newblock \enquote{Bayesian Checking of the Second Levels of Hierarchical
  Models.}
\newblock {\em Statistical Science\/}, 22: 322--343.
\endbibitem

\bibitem[{Bickel(2018)}]{bickel18}
Bickel, D.~R. (2018).
\newblock \enquote{Bayesian revision of a prior given prior-data conflict,
  expert opinion, or a similar insight: a large-deviation approach.}
\newblock {\em Statistics\/}, 52: 552--570.
\endbibitem

\bibitem[{Bousquet(2008)}]{bousquet08}
Bousquet, N. (2008).
\newblock \enquote{Diagnostics of prior-data agreement in applied {B}ayesian
  analysis.}
\newblock {\em Journal of Applied Statisics\/}, 35: 1011--1029.
\endbibitem

\bibitem[{Box(1980)}]{box80}
Box, G. E.~P. (1980).
\newblock \enquote{{Sampling and Bayes' inference in scientific modelling and
  robustness (with discussion)}.}
\newblock {\em Journal of the Royal Statistical Society, Series A\/}, 143:
  383--430.
\endbibitem

\bibitem[{Capp{\'e} et~al.(2005)Capp{\'e}, Moulines, and Ryden}]{cappe+mr05}
Capp{\'e}, O., Moulines, E., and Ryden, T. (2005).
\newblock {\em Inference in Hidden Markov Models (Springer Series in
  Statistics)\/}.
\newblock Secaucus, NJ, USA: Springer-Verlag New York, Inc.
\endbibitem

\bibitem[{Carvalho et~al.(2009)Carvalho, Polson, and Scott}]{carvalho+ps09}
Carvalho, C.~M., Polson, N.~G., and Scott, J.~G. (2009).
\newblock \enquote{Handling Sparsity via the Horseshoe.}
\newblock In van Dyk, D. and Welling, M. (eds.), {\em Proceedings of the Twelth
  International Conference on Artificial Intelligence and Statistics\/},
  volume~5 of {\em Proceedings of Machine Learning Research\/}, 73--80. Hilton
  Clearwater Beach Resort, Clearwater Beach, Florida USA: PMLR.
\endbibitem

\bibitem[{Carvalho et~al.(2010)Carvalho, Polson, and Scott}]{carvalho+ps10}
--- (2010).
\newblock \enquote{The horseshoe estimator for sparse signals.}
\newblock {\em Biometrika\/}, 97(2): 465--480.
\endbibitem

\bibitem[{Clarke and Gustafson(1998)}]{clarke+g98}
Clarke, B. and Gustafson, P. (1998).
\newblock \enquote{On the overall sensitivity of the posterior distribution to
  its inputs.}
\newblock {\em Journal of Statistical Planning and Inference\/}, 71: 137--150.
\endbibitem

\bibitem[{Dahl et~al.(2007)Dahl, G\r{a}semyr, and Natvig}]{dahl+gn07}
Dahl, F.~A., G\r{a}semyr, J., and Natvig, B. (2007).
\newblock \enquote{A robust conflict measure of inconsistencies in {B}ayesian
  hierarchical models.}
\newblock {\em Scandinavian Journal of Statistics\/}, 34: 816--828.
\endbibitem

\bibitem[{Dey et~al.(1998)Dey, Gelfand, Swartz, and Vlachos}]{dey+gsv98}
Dey, D.~K., Gelfand, A.~E., Swartz, T.~B., and Vlachos, P.~K. (1998).
\newblock \enquote{A simulation-intensive approach for checking hierarchical
  models.}
\newblock {\em Test\/}, 7: 325--346.
\endbibitem

\bibitem[{Draper(1995)}]{draper95}
Draper, D. (1995).
\newblock \enquote{Assessment and propagation of model uncertainty (with
  discussion).}
\newblock {\em Journal of the Royal Statisical Society, Series B\/}, 57:
  45--70.
\endbibitem

\bibitem[{Evans(2015)}]{evans15}
Evans, M. (2015).
\newblock {\em Measuring Statistical Evidence Using Relative Belief\/}.
\newblock Taylor \& Francis.
\endbibitem

\bibitem[{Evans and Jang(2010)}]{evans+j10}
Evans, M. and Jang, G.~H. (2010).
\newblock \enquote{Invariant P-values for model checking.}
\newblock {\em The Annals of Statistics\/}, 38: 512--525.
\endbibitem

\bibitem[{Evans and Jang(2011{\natexlab{a}})}]{evans+j11b}
--- (2011{\natexlab{a}}).
\newblock \enquote{A limit result for the prior predictive applied to checking
  for prior-data conflict.}
\newblock {\em Statistics and Probability Letters\/}, 81(8): 1034 -- 1038.
\endbibitem

\bibitem[{Evans and Jang(2011{\natexlab{b}})}]{evans+j11}
--- (2011{\natexlab{b}}).
\newblock \enquote{Weak Informativity and the Information in One Prior Relative
  to Another.}
\newblock {\em Statistical Science\/}, 26: 423--439.
\endbibitem

\bibitem[{Evans and Moshonov(2006)}]{evans+m06}
Evans, M. and Moshonov, H. (2006).
\newblock \enquote{Checking for prior-data conflict.}
\newblock {\em Bayesian Analysis\/}, 1: 893--914.
\endbibitem

\bibitem[{Fan and Lv(2018)}]{fan+lv18}
Fan, J. and Lv, J. (2018).
\newblock \enquote{Sure Independence Screening.}
\newblock In {\em Wiley StatsRef: Statistics Reference Online\/}.
  doi:10.1002/9781118445112.stat08043, Wiley.
\endbibitem

\bibitem[{Gelman et~al.(1996)Gelman, Meng, and Stern}]{gelman+ms96}
Gelman, A., Meng, X.-L., and Stern, H. (1996).
\newblock \enquote{Posterior predictive assessment of model fitness via
  realized discrepancies.}
\newblock {\em Statistica Sinica\/}, 6: 733--807.
\endbibitem

\bibitem[{G\r{a}semyr and Natvig(2009)}]{gasemyr+n09}
G\r{a}semyr, J. and Natvig, B. (2009).
\newblock \enquote{Extensions of a conflict measure of inconsistencies in
  {B}ayesian hierarchical models.}
\newblock {\em Scandinavian Journal of Statistics\/}, 36: 822--838.
\endbibitem

\bibitem[{Griffin and Hoff(2019)}]{griffin+h17}
Griffin, M. and Hoff, P.~D. (2019).
\newblock \enquote{Testing sparsity-inducing penalties.}
\newblock {\em Journal of Computational and Graphical Statistics\/}, To appear.
\endbibitem

\bibitem[{Gustafson and Clarke(2004)}]{gustafson+c04}
Gustafson, P. and Clarke, B. (2004).
\newblock \enquote{Decomposing posterior variance.}
\newblock {\em Journal of Statistical Planning and Inference\/}, 119(2): 311 --
  327.
\endbibitem

\bibitem[{Held and Sauter(2017)}]{held+s17}
Held, L. and Sauter, R. (2017).
\newblock \enquote{Adaptive prior weighting in generalized regression.}
\newblock {\em Biometrics\/}, 73(1): 242--251.
\endbibitem

\bibitem[{Lavine(1991)}]{lavine91}
Lavine, M. (1991).
\newblock \enquote{Sensitivity in {B}ayesian Statistics: The Prior and the
  Likelihood.}
\newblock {\em Journal of the American Statistical Association\/}, 86(414):
  396--399.
\endbibitem

\bibitem[{Len et~al.(2018)Len, Dai, Englert, and Krivitsky}]{len+ek17}
Len, Y.~L., Dai, J., Englert, B.-G., and Krivitsky, L.~A. (2018).
\newblock \enquote{Unambiguous path discrimination in a two-path
  interferometer.}
\newblock {\em Phys. Rev. A\/}, 98: 022110.
\endbibitem

\bibitem[{Li et~al.(2016)Li, Shang, Ng, and Englert}]{xikun16}
Li, X., Shang, J., Ng, H.~K., and Englert, B.-G. (2016).
\newblock \enquote{Optimal error intervals for properties of the quantum
  state.}
\newblock {\em Phys. Rev. A\/}, 94: 062112.
\endbibitem

\bibitem[{Marshall and Spiegelhalter(2007)}]{marshall+s07}
Marshall, E.~C. and Spiegelhalter, D.~J. (2007).
\newblock \enquote{Identifying outliers in {B}ayesian hierarchical models: a
  simulation-based approach.}
\newblock {\em Bayesian Analysis\/}, 2: 409--444.
\endbibitem

\bibitem[{Nott et~al.(2016)Nott, Wang, Evans, and Englert}]{nott+wee16}
Nott, D.~J., Wang, X., Evans, M., and Englert, B.-G. (2016).
\newblock \enquote{{Checking for prior-data conflict using prior-to-posterior
  divergences.}}
\newblock {\em Statistical Scicence\/}, To appear.
\endbibitem

\bibitem[{O'Hagan(2003)}]{ohagan03}
O'Hagan, A. (2003).
\newblock \enquote{{HSS} model criticism (with discussion).}
\newblock In Green, P.~J., Hjort, N.~L., and Richardson, S.~T. (eds.), {\em
  Highly Structured Stochastic Systems\/}, 423--453. Oxford University Press.
\endbibitem

\bibitem[{Paris and \v{R}eh\'a\v{c}ek(2004)}]{paris+r04}
Paris, M. and \v{R}eh\'a\v{c}ek, J. (eds.) (2004).
\newblock {\em Quantum State Estimation\/}, volume 649 of {\em Lecture Notes in
  Physics\/}.
\newblock Springer-Verlag Berlin Heidelberg, 1st edition.
\endbibitem

\bibitem[{Presanis et~al.(2013)Presanis, Ohlssen, Spiegelhalter, and
  Angelis}]{presanis+osd13}
Presanis, A.~M., Ohlssen, D., Spiegelhalter, D.~J., and Angelis, D.~D. (2013).
\newblock \enquote{Conflict Diagnostics in Directed Acyclic Graphs, with
  Applications in {B}ayesian Evidence Synthesis.}
\newblock {\em Statistical Science\/}, 28: 376--397.
\endbibitem

\bibitem[{{Reimherr} et~al.(2014){Reimherr}, {Meng}, and
  {Nicolae}}]{reimherr+mn14}
{Reimherr}, M., {Meng}, X.-L., and {Nicolae}, D.~L. (2014).
\newblock \enquote{{Being an informed Bayesian: Assessing prior informativeness
  and prior likelihood conflict. arXiv1406.5958}.}
\newline\urlprefix\url{http://arxiv.org/abs/1406.5958}
\endbibitem

\bibitem[{R\'{e}nyi(1961)}]{renyi61}
R\'{e}nyi, A. (1961).
\newblock \enquote{On Measures of Entropy and Information.}
\newblock In {\em Proceedings of the Fourth Berkeley Symposium on Mathematical
  Statistics and Probability, Volume 1: Contributions to the Theory of
  Statistics\/}, 547--561. Berkeley, Calif.: University of California Press.
\endbibitem

\bibitem[{Roos et~al.(2015)Roos, Martins, Held, and Rue}]{roos+mhr15}
Roos, M., Martins, T.~G., Held, L., and Rue, H. (2015).
\newblock \enquote{Sensitivity Analysis for Bayesian Hierarchical Models.}
\newblock {\em Bayesian Analysis\/}, 10: 321--349.
\endbibitem

\bibitem[{Scheel et~al.(2011)Scheel, Green, and Rougier}]{scheel+gr11}
Scheel, I., Green, P.~J., and Rougier, J.~C. (2011).
\newblock \enquote{A Graphical Diagnostic for Identifying Influential Model
  Choices in {B}ayesian Hierarchical Models.}
\newblock {\em Scandinavian Journal of Statistics\/}, 38(3): 529--550.
\endbibitem

\bibitem[{Seth et~al.(2019)Seth, Murray, and Williams}]{seth+mw19}
Seth, S., Murray, I., and Williams, C. K.~I. (2019).
\newblock \enquote{Model Criticism in Latent Space.}
\newblock {\em Bayesian Analysis\/}, 14(3): 703--725.
\endbibitem

\bibitem[{Shang et~al.(2013)Shang, Ng, Sehrawat, Li, and Englert}]{shang13}
Shang, J., Ng, H.~K., Sehrawat, A., Li, X., and Englert, B.-G. (2013).
\newblock \enquote{Optimal error regions for quantum state estimation.}
\newblock {\em New Journal of Physics\/}, 15(12): 123026.
\endbibitem

\bibitem[{Teo(2015)}]{teo15}
Teo, Y.~S. (2015).
\newblock {\em Introduction to Quantum-State Estimation\/}.
\newblock Singapore: World Scientific.
\endbibitem

\bibitem[{Tibshirani(1996)}]{tibshirani96}
Tibshirani, R. (1996).
\newblock \enquote{Regression Shrinkage and Selection via the Lasso.}
\newblock {\em Journal of the Royal Statistical Society, Series B\/}, 58:
  267--288.
\endbibitem

\bibitem[{Yuan and Johnson(2012)}]{yuan+j12}
Yuan, Y. and Johnson, V.~E. (2012).
\newblock \enquote{Goodness-of-Fit Diagnostics for {B}ayesian Hierarchical
  Models.}
\newblock {\em Biometrics\/}, 68(1): 156--164.
\endbibitem

\bibitem[{Zhu et~al.(2011)Zhu, Ibrahim, and Tang}]{zhu+it11}
Zhu, H., Ibrahim, J.~G., and Tang, N. (2011).
\newblock \enquote{{Bayesian influence analysis: a geometric approach}.}
\newblock {\em Biometrika\/}, 98(2): 307--323.
\endbibitem

\end{thebibliography}


\begin{acknowledgement}
This work is funded by the Singapore Ministry of Education and the
National Research Foundation of Singapore.
David Nott was supported by a Singapore Ministry of Education Academic
Research Fund Tier 1 grant (R-155-000-189-114).
Hui Khoon Ng is funded by a Yale-NUS College start-up grant.
Michael Evans was supported by a grant from the Natural Sciences 
and Engineering Research Council of Canada.
\end{acknowledgement}

\end{document}